\documentclass[10pt]{article}

\usepackage{graphicx}
\usepackage[numbers]{natbib}
\usepackage{algorithm}
\usepackage{algorithmic}

\usepackage{verbatim}
\usepackage{mathrsfs}

\usepackage[utf8]{inputenc}
\usepackage{amssymb,amsthm,amsmath}
\usepackage{multirow}
\usepackage{url}

\newtheorem{example}{Example}

\begin{document}

\title{A New In-Situ Combustion Simulator for Parallel Computers}

{\author{Ruijian He, Bo Yang, Hui Liu, Zhangxin Chen
		\\
		\vspace{6pt} \\
		Dept. of Chemical and Petroleum Engineering, University of Calgary\\
		2500 University Drive NW, Calgary, AB, Canada\\
	}
}

\date{}
\maketitle

\begin{abstract}
    As a competitive recovery method for heavy oil, In-Situ Combustion (ISC) shows its great potential
    accompanied by technological advances in recent years. Reservoir simulation will play an indispensable
    role in the prediction of the implementation of ISC projects. With the computational complexity, it is
    imperative to develop an effective and robust parallel in-situ combustion simulator.
    In this paper, a mathematical model for In Situ Combustion is proposed, which takes full consideration
    for related physical phenomena, including multi-dimensional multi-component three-phase flow, heat
    convection and conduction, chemical reactions, and mass transfer between phases.
    In the mathematical
    model, different governing equations and constraints are involved, forming a complicated PDE (partial
    differential equation) system. For physical and chemical behaviors, some special treatments for the ISC
    simulator are discussed and applied. Also, a modified PER (Pseudo-Equilibrium Ratio) method is proposed in
    the thesis.
    A fully implicit scheme is applied, and discretization is implemented with the
    FDM (Finite Difference Method). In solving nonlinear systems, the Newton Method is introduced, and both
    numerical and analytical Jacobian matrices are applied. Due to the complexity of an ISC problem, an
    appropriate decoupling method must be considered. Thus the Gauss-Jordan transformation is raised. Then,
    with certain preconditioners and iterative solvers, a numerical solution can be obtained.
    The results of different models are given, which are validated with the results from CMG STARS. Also, the
    scalability of parallelization is proved, indicating the excellent performance of parallel computing. This
    accurate, efficient, parallel ISC simulator applies to complex reservoir models.
\end{abstract}

\section{Introduction}

The International Energy Agency reports that global demand for oil will rise from 90 million barrels per day
in 2012 to 121 million barrels per day in 2040. The main driver of the increase will be China and India
(\cite{conti2016international}). However, reserves of conventional crude oil are declining, and the emphasis
is shifting from conventional oil production to the development of unconventional sources, including oil
sands. Canada’s proven petroleum hydrocarbon reserves include the oil sands in the Athabasca Basin in
northeastern Alberta and Saskatchewan. These deposits, regarded as a safe and secure source of oil for North
America, are the second largest in the world (\cite{kelly2010oil}). However, these oil sands resources have
extremely high viscosity and cannot flow under the reservoir conditions. To reduce its viscosity, thermal
recovery methods, for example, Steam Flooding, Cyclic Steam Stimulation (CSS), Steam Assisted Gravity Drainage
(SAGD) and In-Situ Combustion (ISC), are widely used in reservoirs in Canada.

In-Situ Combustion was first discovered in the 1900s and has been researched for decades. Heat generated
in-situ by oxidation of hydrocarbons in a reservoir can effectively reduce the viscosity of bitumen, which
makes ISC highly energy efficient and productive with outstanding recovery rates, thus becoming attractive
especially to deep or thin reservoirs. However, field applications of ISC have not yielded as much success in
Canada as compared to other thermal methods such as SAGD. It is considered that the challenge of controlling
and predicting the process of ISC is the key obstacle.
In contrast to the steam injection method, steam in the ISC process is generated in the reservoir. An
oxygen-containing gas such as air is injected into the reservoir. And ignition is done near the wellbore,
forming an oxidation zone, where oxidation reactions occur between hydrocarbons and oxygen. Heat generated
from the reactions raised the temperature of a reservoir adjusted to an oxidation zone, where the viscosity of
heavy oil is reduced. Also, steam and gases produced by the reactions, besides the propagation of a combustion
front, displaced oil near the oxidation zone toward a producing well (\cite{moore1995situ, mahinpey2007situ,
ursenbach2010air, chen2014kinetic}).

Reservoir simulation is an area of reservoir engineering in which computer models are used to predict the flow
of fluids (typically, oil, water, and gas) through porous media. The purpose of the simulation is to estimate
field performance, for example, oil recovery under one or more producing schemes.
Many attempts were made for the early mathematical models. In the late 1950s, conduction was assumed to be the
only one mode of heat transfer. And in 1960, convection was also included, but heat losses to the strata were
neglected (\cite{ramey1959transient, bailey1959heat, bailey1960conduction}). Chu proposed a numerical model
for the radial flow of combustion with conduction, convection and heat loss (\cite{chu1963two}). Thomas also
described an analytical treatment of combustion in an infinite system (\cite{thomas1963study}).  Since
computers were introduced to the oil industry in the 1950s, engineers began to leverage mathematical models in
performing reservoir engineering calculations.

Gottfried presented a linear numerical model which described the thermal recovery of 
oil\cite{gottfried1965mathematical}.
James and Farouq Ali developed a 2D mathematical model of the forward combustion process \cite{smith1971simulation}.
Alder presented a new mathematical model of in-situ combustion as a further development of Gottfried’s model.
The model included physical-chemical processes, including combustion of the oil components both in the oil
phase and the gas phase, as well as the effect on the viscosities and heat transfer \cite{adler1975linear}.
Farouq Ali developed a 2D three-phase four-component (one oil component) simulator for 
in-situ combustion\cite{ali1977multiphase}.
Crookston et al. presented a semi-implicit two-dimensional three-phase-flow model with
vaporization-condensation effects \cite{crookston1979numerical}.
Rubin and Vinsome developed a comprehensive one-dimensional in-situ combustion simulator, which was highly
implicit and stable, providing an option of the sequential implicit scheme that made it possible to extend to
a two- or three-dimensional model\cite{rubin1980simulation}.
Grabowski et al. introduced the sequential implicit method into a thermal model, describing both in-situ
combustion and steam processes\cite{grabowski1979fully}.
Youngren described a 3D three-phase in-situ combustion simulator which factored in five components: water,
oxygen, two arbitrary volatile components distributed between the oil and gas phases, and non-volatile oil for
combustion\cite{youngren1980development}.
Coats put forward a general fully-implicit four-phase multi-component multi-dimensional combustion model.
Hwang, Jines, and Odeh developed a new black-oil simulator for an ISC process named 
THERMS\cite{hwang1982situ,odeh1969reservoir}.
Sensitivity studies were made for the in-situ combustion simulators. Anis et al. found that flow convection
parameters, such as a grid block size and dimensionality, were the crucial factors. Numerical distortion of
the convection effects with large grid blocks limits the utility of such models (\cite{anis1983sensitivity}).
Lin et al. used Crookston’s model and found that simulation results were always sensitive to the K-values of a
light oil component. Therefore, more than two crude oil components were required to simulate the evaporation
effect of crude oil accurately (\cite{lin1984numerical}).
Rubin and Buchanan improved the previous model by introducing a fully coupled implicit wellbore model, which
eliminated the lagging of a rate or pressure between the wellbore and the reservoir\cite{rubin1985general}.
Ito and Chow focused on the field-scale simulation of ISC and developed a new and numerically stable algorithm
to achieve the desired fuel consumption\cite{ito1988field}.
Oklany noticed that the in-situ combustion models in the past were private and he developed a model accessible
to the academic community\cite{oklany1992situ}.

Traditionally, kinetic upscaling has been applied to field scale ISC simulation. It is well known that the
kinetics developed from laboratory scale experiment simulation cannot be directly used for field scale
simulation to produce a reliable prediction. The Arrhenius equation shows that the reaction rate depends on
temperature and time through the fuel dependence. The dependence on temperature makes the reactions sensitive
to a grid size because large grid blocks need more heat to achieve a certain temperature level
(\cite{gutierrez2012abcs}). Several attempts have been made to upscale kinetics for field scale simulation. In
some cases, the activation energy was adjusted to a low level and, therefore, the oil burned more easily
(\cite{coats1983some}).
Zhu et al. proposed a new upscaling method with non-Arrhenius kinetics\cite{zhu2011efficient,zhu2011upscaling}.
Nissen et al. further extended previous work and proposed the Work-Flow-Based Upscaling-for-Grid-Independence
(WUGI) scheme, in which Arrhenius kinetics was replaced by equivalent sources terms to represent heat
generation and fuel consumption (\cite{nissen2015upscaling}).
Akkutlu and Yortsos proposed a different approach for ISC simulation with an analytical framework for an ISC
combustion front\cite{akkutlu2003dynamics}.
Kristensen designed two ISC models: the Virtual Combustion Tube (VCT) and the Virtual Kinetic 
Cell (VKC)\cite{kristensen2008development}.

Because of the complexity of the physical process, the in-situ combustion simulators are computationally
complex. Parallel supercomputers become more and more computational powerful in the past years. With the
parallelization, the power of a simulator for the in-situ combustion process will increase significantly.
However, ISC simulators on supercomputers, especially on shared-memory parallel computers, which can run on
thousands of CPUs, are not currently available yet. The primary objective of this study is to design a capable
and robust parallel In-Situ Combustion simulator.
The model in the simulator is adopted from historical research. However, some modifications are required for a
general model. Also, the simplicity of parallelization should also be considered. For the phase equilibrium, a
modified PER (Pseudo-Equilibrium Ratios) method is purposed, which will prevent a singular phenomenon of an
ISC problem. Besides, in the solution process of the ISC simulator, traditional decoupling methods are not so
powerful. A new decoupling method is proposed especially for this complex problem. With this method, the
equation alignment problem is also solved.  As a result, the simulator with the improvements above illustrates
its effectiveness. The accuracy is expected to be shown in the validation with the results from CMG STARS.
Also, the profiles of some variables are shown to coincide with their physical phenomena. In parallelization,
scalability and parallelization should be guaranteed in the simulator.

The structure of this paper is as follows. In \S \ref{sec-math}, the model is introduced. 
The numerical methods are presented in \S \ref{sec-num}, and in \S \ref{sec-results}, numerical studies are
provided to study the implementation and the performance of the parallel simulator.

\section{Mathematical Model}
\label{sec-math}

The mathematical model and governing equations for in-situ combustion are introduced.
For the ISC process, the mathematical model is a system of nonlinear partial differential equations describing multi-phase multi-component fluid flow and heat transfer. Chemical reactions and pressure-volume-temperature (PVT) relationships of the reservoir fluids are all included.

\subsection{Mathematical Description}
Several characteristics of physical and chemical processes are described, including:
\begin{itemize}
  \item Viscosity, gravity, and capillary forces;
  \item Heat conduction and convection processes;
  \item Heat losses to overburden and underburden of a reservoir;
  \item Mass transfer between phases;
  \item Effects of temperature on the physical property parameters of oil, gas, and water;
  \item Chemical reactions, resulting in a change in mass of certain components and temperature;
  \item Rock compression and expansion.
\end{itemize}

These factors can all be reflected in different governing equations, constraints and calculations of properties. With all equations and properties combined, the In-situ Combustion process can be well described.

\subsubsection{Darcy’s Law}

The velocity of a fluid phase is the key to the flow of that fluid, which can measure material and energy spatial exchange. In porous media, permeabilities of different phases are different, where effective and relative permeabilities are introduced. In addition, the pressure and viscosity of different phases differ. Therefore, th velocity of each phase is considered separately by Darcy’s law. The solid phase, representing the existence of coke, does not flow in porous media. There are three flow equations for the gas phase, the oil phase and the water phase, with a lower case denoting each phase with $g$, $o$, $w$ separately (\cite{chen2007reservoir}).
\begin{gather}
\begin{split}
\vec{u}_w & = -\frac{k_{rw}}{\mu_w} \vec{k} \left( \nabla p_w - \gamma_w \nabla z\right)\\
\vec{u}_o & = -\frac{k_{ro}}{\mu_o} \vec{k} \left( \nabla p_o - \gamma_o \nabla z\right)\\
\vec{u}_g & = -\frac{k_{rg}}{\mu_g} \vec{k} \left( \nabla p_g - \gamma_g \nabla z\right)
\end{split}
\label{eq4.1}
\end{gather}
In the above equations, $\gamma_\alpha, \alpha = w,o,g$, is the specific weight of a certain phase, which is computed by the phase density $\rho_\alpha$ and acceleration due to gravity $g$. $\mu_\alpha$ and $k_{r\alpha}$ denote the dynamic viscosity and relative permeability of a phase.
Darcy's law applies to laminar flow, where the fluid movement is dominated by viscous forces. This occurs when a fluid moves slowly, along parallel streamlines. When a flow velocity increases or a kinetic viscosity decreases, a fluid moves chaotically and the streamlines are no longer in parallel formation. Flow acts in turbulent flow and the inertial forces are more influent than the viscous forces.

\subsubsection{Mass Conservation Equations}

For a multi-phase, multi-component system, we introduce the molar fraction $x_{c,\alpha}$ of a component in the $\alpha$-phase. The molar number of a component in a phase and the total molar number of the phase are denoted as $n_{c,\alpha}$ and $n_{\alpha}$, respectively. Thus the molar fractions are
\begin{gather}
x_{c,\alpha} = \frac{n_{c,\alpha}}{n_{\alpha}}.
\end{gather}
The total molar number of component $c$ is conserved as below (\cite{chen2007reservoir}):
\begin{gather}
\frac{\partial}{\partial t}\left(\phi \Sigma_{\alpha}^{N_\alpha} \rho_\alpha S_\alpha x_{c,\alpha}\right) =
- \nabla \cdot \left( \Sigma_{\alpha}^{N_\alpha} \rho_\alpha S_\alpha \vec{u}_{\alpha} \right)
+ \Sigma_{\alpha}^{N_\alpha} q_{\alpha,well} x_{c,\alpha} + q_{c, reac}.
\label{eq4.2}
\end{gather}
Also, we have a single mass conservation for coke (\cite{oklany1992situ}):
\begin{gather}
\frac{\partial}{\partial t}C_c =q_{c, reac}.
\label{eq4.8}
\end{gather}
Since the solid phase does not flow, there is no convection term in the equation. Only chemical reaction terms can affect the coke concentration.

In this equation, it is noticeable that different from other models, the “mass” conserved here is only the molar number rather than the mass. Also, $\rho_\alpha$ and $q_\alpha$ are the molar density and molar production/injection of phase $\alpha$. Specifically, $q_{c, reac}$ here represents a reaction term, which has a similar effect as a well term. When a component is the reactant of a reaction, the sign of the reaction term can be negative, since the molar number of that component will decrease when the reaction takes place. Also, if a component is the product of a reaction, the sign can be positive due to an increase in its molar number during the reaction. This reaction term can be a sum of several reactions, since a component can be a reactant or a product in different reactions. For example, oxygen is consumed in each oxidation reaction with different rates, so this term can be a superposition.

\subsubsection{Energy Conservation Equation}

The energy conservation equation must be taken into account besides the mass conservation equation for a thermal process (\cite{chen2007reservoir}):
\begin{gather}
\begin{split}
& \frac{\partial}{\partial t}\left(\phi(\rho_w S_w U_w + \rho_o S_o U_o + \rho_g S_g U_g) + (1-\phi)U_r + C_c U_c\right) \\
= \quad & \nabla \cdot \left( K_T \nabla T \right)
- \nabla \cdot \left( \rho_w H_w \vec{u}_w + \rho_o H_o \vec{u}_o + \rho_g H_g \vec{u}_g \right)\\
& + (q_{w,well} H_w + q_{o,well} H_o + q_{g,well} H_g) + Q_{reac} - Q_{loss}.
\end{split}
\label{eq4.3}
\end{gather}
In the equation above, the only causes for an energy change are heat conduction, heat convection, and kinetic
and potential energy. The effects of a viscous force and an energy addition due to radiation and
electromagnetism are ignored.  On the left-hand side of the equation, the variations in the bracket represent
the molar internal energy of all fluid phases and the solid phase (which only contains coke as a component),
as well as the rock. $U$ denotes the volumetric internal energy.  On the right-hand side, the first term
represents the conduction term. This is caused by a difference in temperature, where the rate of conduction is
constraint by $K_T$, the bulk thermal conductivity. The thermal conductivity here is a combination of liquid,
rock and solid matter, where a linear mixing rule is applied (\cite{cmg2015starguide}). For the ISC process,
coke is considered in the normal mixing rule:
\begin{gather}
K_T = \phi \left[S_w K_w + S_o K_o + S_g K_g \right] + (1-\phi)K_r + C_c K_c.
\end{gather}
In the equation, $K_w, K_o, K_g, K_r, K_c$ denote thermal conductivities for water phase, oil phase, gas
phase, rock and coke separately.

The second term denotes the heat convection, a result of fluid flow. The molar enthalpy of different phases
are different and an enthalpy change occurs when a fluid flows. The third term represents the energy exchange
rate due to a source/sink well, which is negative for production and positive for injection. The expression is
similar to the convection term, since both describe an enthalpy change due to fluid flow. For the fourth term,
a thermal effect of reactions is described, which can be a positive term (exothermic reaction) or a negative
term (endothermic reaction). 

A heat loss term to underburden and overburden is also considered in the last term (\cite{chen2007reservoir}):
\begin{gather}
Q_{loss} = A K_{ob} \left( \frac{T(ob)-T_{ini}}{d} - \rho \right),
\end{gather}
where $A$ is the interface area with overburden/underburden, $T(ob)$ is the temperature at the interface, and
$\rho$ is a fitted parameter.

\subsubsection{Chemical Reaction}

For an ISC simulator, the main difference between the ISC simulator and a pure thermal recovery simulator is the involvement of chemical reactions. As mentioned above, reaction terms are added to the mass conservation equations and energy conservation equation. For the mass conservation equation of component $c$, the reaction term $q_{c,reac}$ can be expressed as below (\cite{chen2007reservoir}):
\begin{gather}
q_{c,reac} = \sum_{R}^{N_R} {s_{c,R}r_R},
\end{gather}
where $r_R$ is the reaction rate of reaction $R$ and $s_{c,R}$ is the stoichiometric coefficient of component $c$ in reaction $R$. With a stoichiometric coefficient, if component $c$ is a reactant, then we note that this coefficient is negative. If a component is a product, then it will be taken as a positive number. If it does not participate in a reaction, the coefficient will then be zero. With the stoichiometric coefficients, the chemical reaction equation can be expressed as follows:
\begin{gather}
\sum_{c}^{N_c}{s_{c,R} \cdot (\mbox{Chemical formula of component } c)} \longrightarrow 0.
\end{gather}
With all the stoichiometry matrix listed together, the following stoichiometry matrix can be obtained:
\begin{gather}
 N = (s_{c,R})_{c,R} = \left[
 \begin{matrix}
   s_{1,1} & s_{2,1} & ... & s_{N_c, 1} \\
   s_{1,2} & s_{2,2} & ... & s_{N_c, 2} \\
   ...     & ...     & ... & ...        \\
   s_{1,N_R} & s_{2,N_R} & ... & s_{N_c, N_R}
  \end{matrix}
  \right].
\end{gather}

When a chemical reaction model is input into our simulator, all the coefficients must be checked. Otherwise,
mass conservation of the total system may not hold. For chemical reactions that obey the law of conservation
of matter, the following equation holds:
\begin{gather}
N \vec{M} = \vec{0},
\end{gather}
where $\vec{M}$ is the vector of molar masses of components. The c-th entry of the vector is the molar mass of
component $c$.

As for the reaction term in the energy conservation equation, it can be expressed as:
\begin{gather}
Q_{reac} = \sum_{R}^{N_R} {H_{R}r_R},
\end{gather}
where $H_R$ is the reaction enthalpy of the $R$-th reaction (\cite{coats1980situ}).  Expressions of reaction
terms in both the mass conservation equations and energy conservation equation are all in the form of a linear
combination of reaction rates. Therefore, calculations of a reaction rate are crucial in the ISC simulator.
Furthermore, some constraint equations are required to solve these conservation equations. Constraint
equations are functions that do not involve any convection or conduction term and are only dependent on local
properties. They express all secondary variables and parameters as functions of a set of primary thermodynamic
variables selected.

\subsubsection{Phase Saturation Constraint}

The saturation constraint forces all the available pores of rock to be saturated with fluids, which is encountered for in all multi-phase flows (\cite{chen2007reservoir}):
\begin{gather}
S_w + S_o + S_g = 1.
\label{eq4.4}
\end{gather}
This constraint indicates that a pore volume is occupied only by the water, oil, and gas phases. Specifically,
we only take these three phases into account as coke does not take a volume obviously, and, therefore, a
saturation for the solid phase is not introduced.

\subsubsection{Capillary Pressure Constraints}

A constraint is a conversion from the oil pressure (the default system pressure) to the water pressure and gas
pressure. A capillary pressure $P_c$ is the pressure difference across the interface between two immiscible
fluids arising from capillary forces. These capillary forces are surface tension and interfacial tension.
Capillary pressures are usually functions of saturation. In a three-phase system, we have the following
relationship (\cite{chen2007reservoir}):
\begin{gather}
p_w = p_o - p_{cow}(S_w), \quad p_g = p_o + p_{cog}(S_g).
\label{eq4.5}
\end{gather}
These capillary pressures are assumed to be known.

\subsubsection{Phase Composition Constraints}

A constraint implies that the sum of all the components' mole fractions in a phase adds up to one, which is
usually encountered for in compositional flow (\cite{chen2007reservoir}):
\begin{gather}
\Sigma_{\alpha}^{N_\alpha} x_{c,\alpha} = 1, \quad \alpha = w,o,g.
\label{eq4.6}
\end{gather}
It comes from the total mole number of a given phase that
\begin{gather}
\Sigma_{\alpha}^{N_\alpha} n_{c,\alpha} = n_{\alpha}, \quad \alpha = w,o,g.
\end{gather}

\subsubsection{Phase Equilibrium Constraints}

A constraint characterizes the variation of mass distribution of a given component in its phases. Based on the
assumption that the mass interchange between phases occurs much faster than the flow in porous media, all the
phases are in the equilibrium state. Consequently, the distribution of each component is subjected to the
condition of a stable thermodynamic equilibrium, which is given by minimizing the Gibbs free energy of the
compositional system. The component distributed phases are arbitrary and determined by the component
properties. The following equation expresses the thermodynamic equilibrium state of component $c$ being
distributed in two different phases $\alpha_1$ and $\alpha_2$ (\cite{chen2007reservoir}):
\begin{gather}
f_{c,\alpha_1}(p_{\alpha_1}, T, x_{1,\alpha_1}, \cdots, x_{N_c,\alpha_1}) = f_{c,\alpha_2}(p_{\alpha_2}, T, x_{1,\alpha_2}, \cdots, x_{N_c,\alpha_2}).
\label{eq4.7}
\end{gather}

In the equilibrium equation, $f_{c,\alpha}$ denotes the fugacity of component $c$ in phase $\alpha$, which is
an effective partial pressure that replaces the mechanical partial pressure in an accurate computation of the
chemical equilibrium constant. It is a function of pressure, temperature, and molar fractions of all
components existing in a phase.  Several mathematical techniques handle the mass interchange in different
phases of chemical components. The most common ones are based on (1) the K value approach, (2) an equation of
state, and (3) a variety of empirical tables from experiments.

In a multi-component system, a K value (or an equilibrium ratio) is defined as the ratio of the mole fractions of a component in its distributed two phases:
\begin{gather}
{K}_{c,\alpha_1,\alpha_2} = \frac{x_{c,\alpha_1}}{x_{c,\alpha_2}}.
\end{gather}
For ideal solutions, a K value is a function of only the system pressure and temperature, regardless of the overall composition of the mixture. It may be calculated from an analytic equation as:
\begin{gather}
K = \left(\frac{{kv}_1}{p} + {kv}_2 p + {kv}_3\right)\exp\left(\frac{{kv}_4}{T-{kv}_5}\right).
\end{gather}
For our simulator, the K value approach is used. Due to the complexity of a non-isothermal process, the
analytic equation is used for phase saturation constraint calculations. However, for a real solution, a K
value is not only a function of pressure and temperature, but also depends on the composition of the system.
While the K value approach is easy to set up, it lacks generality and may result in inaccurate reservoir
simulation. In recent years, an EOS has been more widely employed due to the more consistent compositions,
densities and molar volumes produced. The most famous EOS is the van der Waals EOS (\cite{van1910equation}).
Here we discuss three more accurate and reliable EOS: Peng–Robinson (PR), Redlich–Kwong (RK), and
Redlich–Kwong–Soave (RKS) (\cite{peng1976new}; \cite{redlich1949thermodynamics}; \cite{soave1972equilibrium}).
The main advantage of using an EOS is that the same equation can be used to model the behavior of all phases,
thereby assuring consistency when performing phase equilibria calculations. Introduction of the compositions
into the K value functions, however, adds considerable complexity to the flash computation
(\cite{cmg2015starguide}; \cite{coats1976simulation}).

\subsubsection{Unknowns and Equations}

In our ISC simulator, $3N_c+11$ unknowns need to be solved for. They consist of three pressures, saturations,
velocities for each phase, and molar fraction in each phase of each component. Also, temperature and
concentration of coke, which is the only component of the solid phase, are also involved. As for the
equations, mass conservation, coke conservation, energy conservation and Darcy’s law are included, which add
up to $N_c+5$ equations. The phase equilibrium constraints, phase composition constraints, phase saturation
constraint and capillary pressure constraints add up to $2N_c+6$ equations. Therefore, we have $3N_c+11$
unknowns and $3N_c+11$ equations; this problem can be solved. Initial conditions and closed boundary
conditions are applied.  The table below lists the unknowns and equations.

\begin{table}[!htb]
\centering
\begin{tabular}{|c|c|c|}
\hline
\textbf{Unknowns} & \textbf{Definition} & \textbf{Number}\\\hline
$p_w, p_o, p_g$ & Pressure of 3 phases & $3$ \\\hline
$T$ & Temperature & $1$ \\\hline
$S_w, S_o, S_g$ & Saturation of 3 phases & $3$ \\\hline
$\vec{u}_w, \vec{u}_o, \vec{u}_g$ & Velocity of 3 phases & $3$\\\hline
$x_{c, w},\quad c = 1,2,\cdots,N_c$  & Mole fraction of single component in water phase & $N_c$\\\hline
$x_{c, o},\quad c = 1,2,\cdots,N_c$  & Mole fraction of single component in oil phase & $N_c$\\\hline
$x_{c, g},\quad c = 1,2,\cdots,N_c$  & Mole fraction of single component in gas phase & $N_c$\\\hline
$C_c$ & Coke concentration & $1$\\\hline
\multicolumn{2}{|c|}{\textbf{Summation}} & \textbf{$3N_c + 11$}\\
\hline
\end{tabular}
\caption{Unknowns of the mathematical model}
\label{Tab4.1}
\end{table}

\begin{table}[!htb]
\centering
\begin{tabular}{|c|c|c|}
\hline
\textbf{Equations} & \textbf{Index in the thesis} & \textbf{Number}\\\hline
Mass conservation of $N_c$ components & (\ref{eq4.2}) & $N_c$\\\hline
Energy conservation equation & (\ref{eq4.3}) & $1$\\\hline
Darcy’s law & (\ref{eq4.1}) & $3$\\\hline
Phase equilibrium equations between phase o \& w & (\ref{eq4.7}) & $N_c$\\\hline
Phase equilibrium equations between phase o \& g & (\ref{eq4.7}) & $N_c$\\\hline
Coke conservation equation & (\ref{eq4.8}) & $1$\\\hline
Phase composition constraint for 3 phases & (\ref{eq4.6}) & $3$\\\hline
Capillary pressure constraint & (\ref{eq4.5}) & $2$\\\hline
Phase saturation constraint & (\ref{eq4.4}) & $1$\\\hline
\multicolumn{2}{|c|}{\textbf{Summation}} & \textbf{$3N_c + 11$}\\
\hline
\end{tabular}
\caption{Equations of the mathematical model}
\label{Tab4.2}
\end{table}

Darcy’s law is used to eliminate the velocities of the three phases; K value calculations are used to
eliminate the molar fraction of each component in the water and gas phases, keeping the molar faction of each
component in the oil phase; capillary pressure constraints are used to eliminate pressures in the water and
gas phases, with the oil pressure kept; the phase saturation constraint is used to eliminate the oil
saturation.  In our simulator, the components we consider can be divided into the water components, oil
components, and gas components. The oil components can only exist in the oil phase and gas phase, and the
water components can only exist in the water phase and gas phase. However, the gas components here denote
non-condensate gas, which can only exist in the gas phase. With the definition of the three kinds of
components, there is no component which can exist both in the oil phase and water phase. Therefore, K value
equations between the oil and water phases are no longer needed, meaning that the number of equations should
decrease by $N_c$.
For the water components, usually we have only one water component:
$$x_{w,w}=1,$$
which means that $N_c$ unknowns are no longer needed. After the elimination, only $N_c+3$ unknowns and
equations are left, forming our final system of unknowns and equations.

\begin{table}[!htb]
\centering
\begin{tabular}{|c|c|c|c|}
\hline
\textbf{Unknowns} & \textbf{Index} & \textbf{Unknowns} & \textbf{Index}\\\hline
$p_o$ & $0$ & Mass equation for comp 1 & $1 - 1 = 0$\\\hline
$T$ & $1$ & Mass equation for comp 2 & $2 - 1 = 1$\\\hline
$S_w$ & $2$ & $\cdots$ & \\\hline
$S_g$ & $3$ & Mass equation for comp i & $i-1$\\\hline
$x_{1,o}$ & $1 + 3 = 4$ & $\cdots$ & \\\hline
$\cdots$ & & Mass equation for comp $N_c$ & $N_c - 1$\\\hline
$x_{c,o}$ & $c + 3$ & Energy equation & $N_c$\\\hline
$\cdots$ & & Oil constraint equation & $N_c + 1$\\\hline
$x_{N_c-1,o}$ & $N_c + 2$ & Gas constraint equation & $N_c + 2$\\\hline
$C_c$ & $N_c + 3$ & Coke equation & $N_c + 3$\\
\hline
\end{tabular}
\caption{Unknowns and equations of In-Situ Combustion Simulator}
\label{Tab4.3}
\end{table}

In our simulator, one water component is mandatory. Also, coke is mandatory for the combustion reaction.
Hence, the breakdown of $N_c$ is
\begin{equation}
N_c = n_{cw} + n_{co} + n_{cg} + n_{cc},
\end{equation}
We actually have
\begin{equation}
N_c = 2 + n_{co} + n_{cg}.
\end{equation}
Here, $n_{c\alpha}$ denotes the number of the ``default" components in phase $\alpha$. For example, $n_{cg}$
here denotes the number of non-condensate gas components. For all other components such as the water, light
oil, and heavy oil components, they are not originally from the gas phase. In addition, a large fraction of
each component is not in the gas phase under the standard conditions. The ``gas components", however, will
meet these two requirements. In the ISC model, a gas component mostly refers to a real component or a pseudo
component such as inert gas, oxygen, carbon-monoxide, and carbon-dioxide.

For non-condensate gas components that do not dissolve into the oil phase, their molar fractions in the oil
phase, $x$, will be zero. Therefore, $x$ can no longer be kept as an unknown for them. The molar fraction of
non-condensate gas in the gas phase will be the new unknowns for them.
For a clearer demonstration, different from the general model, the molar fraction of a component in the gas
phase is denoted by $y$; in the oil phase, it is still denoted by $x$. The numbering of a component is also
noted as a subscript.
For an ISC model, a well is a crucial part and well equations must be included. For each well, according to
different well control conditions, well equations can be different. But for one single well, only one unknown
is required ($bhp$), and other parts of the equation are constants or functions of other unknowns. Therefore,
with the well equations and $bhp$ included, the system is still well-defined.
$p_o$ can be seen as a basic pressure of a reservoir since the pressures of the gas phase and water phase can
be calculated through these capillary pressure functions. Actually, the difference between the oil pressure
and $p_w$, $p_g$ is relatively small in some cases. So we denote $p$ as $p_o$ directly.
At this point, we can obtain a manipulated mathematical system which is shown in Table \ref{Tab:nco&ncg}.

\begin{table}[!htb]
\centering
\begin{tabular}{|c|c|c|}
\hline
\textbf{Index} & \textbf{Unknowns} & \textbf{Equations}\\\hline
0 & $p$ & Water mass equation\\\hline
1 & $x_{0}$ & Oil mass equation for oil component $0$ \\\hline
$\cdots$ & $\cdots$ & $\cdots$\\\hline
$n_{co}$ & $x_{n_{co}-1}$ & Oil mass equation for oil component $n_{co}-1$ \\\hline
$n_{co}$ + 1 & $y_{0}$ & Gas mass equation for gas component $0$ \\\hline
$\cdots$ & $\cdots$ & $\cdots$\\\hline
$n_{co}$ + $n_{cg}$ & $y_{n_{cg}-1}$ & Gas mass equation for gas component $n_{cg}-1$ \\\hline
$n_{co}$ + $n_{cg}$ + 1 & $T$ & Energy equation\\\hline
$n_{co}$ + $n_{cg}$ + 2 & $S_w$ & Oil constraint equation\\\hline
$n_{co}$ + $n_{cg}$ + 3 & $S_g$ & Gas constraint equation\\\hline
$n_{co}$ + $n_{cg}$ + 4 & $C_c$ & Coke equation\\\hline
$n_{co}$ + $n_{cg}$ + 5 & $bhp$ & Well equation\\\hline
\end{tabular}
\caption{Modified mathematical system for ISC}
\label{Tab:nco&ncg}
\end{table}
This is the final table of unknowns and equations. Specifically, the indices of the corresponding unknowns and
equations are also displayed. In the programming of the simulator, these indices are used especially when
building a linear system, including Jacobian matrices and right-hand side vectors in Newton iterations.

\subsection{Properties and Specific Treatments}

\subsubsection{Reaction Model}

To illustrate our general chemical model with relative ease, we introduce a four-phase six-component model as
an example. This model is originally from Crookston’s model (\cite{crookston1979numerical}). The six
components involved in this model are Oxygen, Inert gas, Light oil, Heavy oil, water, and coke. There are four
phases considered in our model. They are Gaseous (gas), Oleic (oil), Aqueous (water) and Solid (coke).
For the gas components, only oxygen and inert gas are considered. The inert gas consists of all the
non-condensate gases except the oxygen, including nitrogen, carbon monoxide, carbon dioxide and other oxides
of carbon. Therefore, the inert gas component is actually a pseudo-component.
Water is assumed to be immiscible with oil. Vaporization and condensation of both water and oil are controlled
by vapor-liquid equilibrium.
The situation of a component existing in different phases can be different on a case-by-case basis. This part
is designed to be user input. The relationship between components and phases is shown in the following table.

\begin{table}[!htb]
\centering
\begin{tabular}{|c|c|c|c|c|c|}
\hline
\multicolumn{2}{|c|}{\multirow{2}*{\textbf{Component}}} & \multicolumn{4}{c|}{\textbf{Phase (Mole Fraction)}} \\\cline{3-6}
\multicolumn{2}{|c|}{~} & \textbf{Gaseous} & \textbf{Oleic} & \textbf{Aqueous} & \textbf{Solid} \\\hline
0 & Water & $y_0 = K_0$ & & $1$ & \\\hline
1 & Light Oil & $y_1 = K_1 x_0$ & $x_0$ & & \\\hline
2 & Heavy Oil & $y_2 = K_2 x_1$ & $x_1$ & & \\\hline
3 & Oxygen & $y_3$ & & & \\\hline
4 & Inert Gas & $y_4$ & & & \\\hline
3 & Coke & & & & 1\\
\hline
\end{tabular}
\caption{Correlation of phase and components in a simple model}
\label{Tab4.4}
\end{table}

For the oil component, we just make a simple breakdown of the light component and heavy component in this
model. Besides the physical properties of the two components, the main difference is that there is a pyrolysis
reaction for the heavy component, but not for the light component. Besides oxidation reactions for the two
components, there are three reactions for oil components.  As the main product of the pyrolysis reaction,
there is an oxidation reaction for coke, which is the main fuel source of heat generation in this model.  As
shown in Table \ref{Tab4.5}, only four reactions are included in this model.

\begin{table}[!htb]
\centering
\begin{tabular}{|c|c|c|c|}
\hline
\textbf{Index} & \textbf{Name} & \textbf{Chemical reaction equation} & \textbf{Reaction rate}\\\hline
1 & Light Oil Oxidation & $LO + s_1 O_2 \longrightarrow s_2 CO_x + s_3 H_2 O$ & $r_1$ \\\hline
2 & Heavy Oil Oxidation & $HO + s_4 O_2 \longrightarrow s_5 CO_x + s_6 H_2 O$ & $r_2$ \\\hline
3 & Heavy Oil Cracking & $HO \longrightarrow s_7 LO + s_8 Coke + s_9 CO_x$ & $r_3$ \\\hline
4 & Coke Oxidation & $Coke + s_{10} O_2 \longrightarrow s_{11} CO_x + s_{12} H_2 O$ & $r_4$ \\
\hline
\end{tabular}
\caption{Unknowns and equations of In-Situ Combustion Simulator}
\label{Tab4.5}
\end{table}

According to the discussion, we know that the reaction sets are directly related to the division of
components. These four reactions are only a chemical reaction model for this certain six-component four-phase
model. If the user inputs a SARA model, with saturated, aromatic, resin and asphaltene as oil components, the
reaction model will be totally different.

From these four reactions, a reactant and production, which are the basic components, can be known with
certainty. Mass change of certain components is described by a kinetic reaction rate, which has a dimension of
$MT^{-1}L^{-3}$. This rate is taken from the Arrhenius equation (\cite{coats1980situ}), a formula describing
the temperature dependence of reaction rates. Arrhenius' equation gives the dependence of the rate constant of
a chemical reaction on the absolute temperature, a pre-exponential factor and other constants of the reaction:
\begin{gather}
k(T) = Ae^{-(E_a/RT)}
\end{gather}
In this equation, $k(T)$ is a reaction rate constant. It is actually a function of temperature, not a
constant; the parameter of concentration is not included here. $E$ is the activation energy of a reaction. The
larger the activation energy, the slower the reaction. $T$ is the reaction temperature. $A$ is a
pre-exponential factor, a constant for each chemical reaction that defines the rate due to the frequency of
collisions in the correct orientation. In fact, this factor is a function of oil type, pressure, and other
parameters.

At a low temperature, a combustion reaction is not likely to happen. Another explanation would be that
oxidation is happening at a very low reaction rate, making the combustion reaction to appear not to happen.
When the temperature is higher, the energy of a molecular system is easier to exceed the activation energy.
Movement of particles becomes much more violent at higher temperature, resulting in more possibility of
effective collisions between particles of reactants. Therefore, the reaction rate will be much higher.
Temperature plays an extremely significant role in chemical reaction rates, and an exponential function
perfectly describes this behavior. A rate will be accelerated exponentially by raising the temperature of a
system.  Thus we have the following accurate expressions of kinetic reaction rates:
\begin{gather}
\begin{split}
r_1 &= A_1 e^{-\frac{E_{a,1}}{RT}} (y_{3} p_g) (\phi S_o \rho_o x_0),\\
r_2 &= A_2 e^{-\frac{E_{a,2}}{RT}} (y_{3} p_g) (\phi S_o \rho_o x_1),\\
r_3 &= A_3 e^{-\frac{E_{a,3}}{RT}} (\phi S_o \rho_o x_{1}) (1 - (\frac{C_c}{C_{c,max}})^5),\\
r_4 &= A_4 e^{-\frac{E_{a,4}}{RT}} (y_{3} p_g) (C_c).
\end{split}
\end{gather}

In these equations, we can see that a reaction rate depends on several physical properties of the reactants, including concentration and pressure. If a reaction occurs between the gas phase and the oil phase, the reaction rate is affected by the concentration of an oil component and partial pressure of a gas component. For a reaction occurring between the solid phase and the gas phase, the rate is proportional to the concentration of coke and partial pressure of the gas component, oxygen. Particularly, for the oil cracking reaction, the term $(1 - (\frac{C_c}{C_{cmax}})^5)$ is added to control the production of coke. When the coke concentration is approaching the concentration limit of coke, this term will reduce the reaction rate, preventing the cracking reaction from producing more coke.
This model is only one fulfillment of the chemical reactions in which oxidation reactions of oil components in
the gas phase are ignored, and oxidation reactions are not divided into the low-temperature and
high-temperature reactions (LTO and HTO). The chemical reactions are open to users of this simulator and any
scheme can be decided and input by the user, such as the SARA model (\cite{belgrave1993comprehensive}).

\subsubsection{Phase Equilibrium}

Due to the complexity of a non-isothermal model, phase equilibrium simulation is really time-consuming, so we use a K value approach to describe the phase behavior. Calculations of K-values are as follows (\cite{chen2007reservoir}; \cite{cmg2015starguide}):

\begin{gather}
\begin{split}
&{K}_{W} = K[0](p, T) \\
= & \left(\frac{kv1_{W}}{p} + kv2_{W} \cdot p + kv3_{W}\right)\exp{\left(\frac{kv4_{W}}{T-kv5_{W}}\right)},\\
&{K}_{O,i} = K[i + 1](p, T) \\
= & \left(\frac{kv1_{O,i}}{p} + kv2_{O,i} \cdot p + kv3_{O,i}\right)\exp{\left(\frac{kv4_{O,i}}{T-kv5_{O,i}}\right)}.\\
\end{split}
\end{gather}
However, the K value calculations above are correct when two phases co-exist: the water component exists in
both the water phase and the gas phase or an oil component exists in both the oil phase and the gas phase.
However, for an ISC process, temperature varies dramatically, leading to significant changes in phases.

For example, when a combustion front propagates, temperature rises rapidly. At first, water will evaporate due
to its lower boiling point compared to the oil phase. With the higher temperature, the oil phase will also
disappear. Since light components evaporate at high temperatures, heavy components will decrease to zero due
to their participation in chemical reactions such as heavy oil cracking and heavy oil oxidation. That is to
say, even if $S_{or}$ is given in a relative permeability table when the saturation of the oil phase is low
enough, mobilization is restricted and the oil phase will also disappear in the ISC process. Ideally, when the
temperature is high enough, at positions near the combustion front, both the water phase and the oil phase
disappear, leading to only the gas phase and solid phase. The water component and oil components take a large
portion of the gas phase. When the combustion process is over, the temperature will decrease. If water or oil
components exist in the gas phase, these gasses will condense, forming the water phase and oil phase.

When such changes in phases occur, the K value formulas or the equation system will not be suitable anymore,
resulting in a failure of the simulator. Therefore, proper handling of the appearance and disappearance of
phases is significant. Typically, we have two approaches to deal with such a problem. First, the VS (Variable
Substitution) method can be applied to the entire system, while the unknowns and equations need to be changed
(\cite{coats1980situ}). Judging the process of the method to be added to the simulator, it would result in
complicated logic in the coding part and discontinuity of the system, leading to instability of the simulator.

As a result, we implement the PER (Pseudo-Equilibrium Ratios) method in our simulator (\cite{crookston1979numerical, abou1985handling}). With this method, a correction term is added to the original K-value, and a pseudo K value is acquired:
\begin{gather}
{K*}_{W} = K^*[0](p, T) = \left(\frac{S_w}{S_w+n_{cg}}\right)K[0](p, T),\\
{K*}_{O,i} = K^*[i + 1](p, T) = \left(\frac{S_o}{S_o+\epsilon}\right) K[i + 1](p, T).
\end{gather}
Therefore, the gas phase molar fraction for the oil components and water component are functions of $p, T, S_w, S_g$. The molar fraction in the gas phase for gas components are the basic unknowns:
\begin{gather}
y = y (p, T, S_w, S_g).
\end{gather}

In calculations of pseudo K-values, $\epsilon$ is a small number of the order of $1e-4$. When the water or oil
phase exists, $S_w \gg \epsilon$, and the term $\left(\frac{S_w}{S_w+\epsilon}\right) \approx 1$. Thus the
pseudo K-values are extremely close to the K-values. As the liquid phase gradually disappears, the saturation
decreases and approaches $\epsilon$, and the term $\left(\frac{S_w}{S_w+\epsilon}\right)$ approaches $0.5$.
When the saturation keeps decreasing, the term $\left(\frac{S_w}{S_w+\epsilon}\right)$ approaches $0$.

For a typical K-value, when the liquid phase disappears, the K value approach cannot be used anymore. We take
pure water as an example. When there is only a water phase, we heat the system and the temperature keeps
rising. Below the boiling point, the K value is always less than 1. When the boiling point at a certain
pressure is achieved, the K value equals 1. This can be used in the calculated saturated steam temperature.
When the temperature of the mixture increases, the water phase disappears, leaving only the gas phase. The K
value under this condition is greater than 1.
Obviously, if the water phase and gas phase co-exist, the molar fraction of the water component in the water
phase and gas phase should be 1, leading to a K value of 1. In other words, a K value is only validated when
it equals 1; i.e., two phases coexist. If the liquid phase and gas phase do not co-exist, the formula of a K
value has no meaning. This is because the K value is defined by formula $K = \frac{y}{x}$, and the liquid
phase and gas phase must appear.

In the ISC process, the disappearance of the liquid phase can be described by the PER method. With an increase
in the K value as temperature rises, the $y$ values can be greater than 1, meaning that the molar fraction of
a certain component is greater than 1. Obviously, this is wrong. With the correction term, however, pseudo
K-values can be less than K-values. Therefore, although phase equilibrium no longer applies, with the pseudo
K-values, the relationship $y = K^* x$ still applies.
When saturation approaches 0, both the pseudo K-values and fraction $y$ of a certain component approach 0.
This process can be described by the formulas perfectly. With the formulas, we can get the saturation by
solving the equation system at each time step. The unknowns and saturation can be calculated to a non-zero
value.
Although its saturation should be zero when the liquid phase disappears, it seems that the PER method
introduces an obvious error deviation in this case. However, the correction term will be effective only when
the liquid phase begins to disappear. At that time, the calculated saturation will be close to $\epsilon$ or
even smaller. So, when we take $\epsilon = 1e-4$, the absolute error of saturation is less than $1e-4$, which
is acceptable, especially when the great convenience the method brings is considered.

\begin{figure}[ht]
 \centering\includegraphics[width=0.75\textwidth]{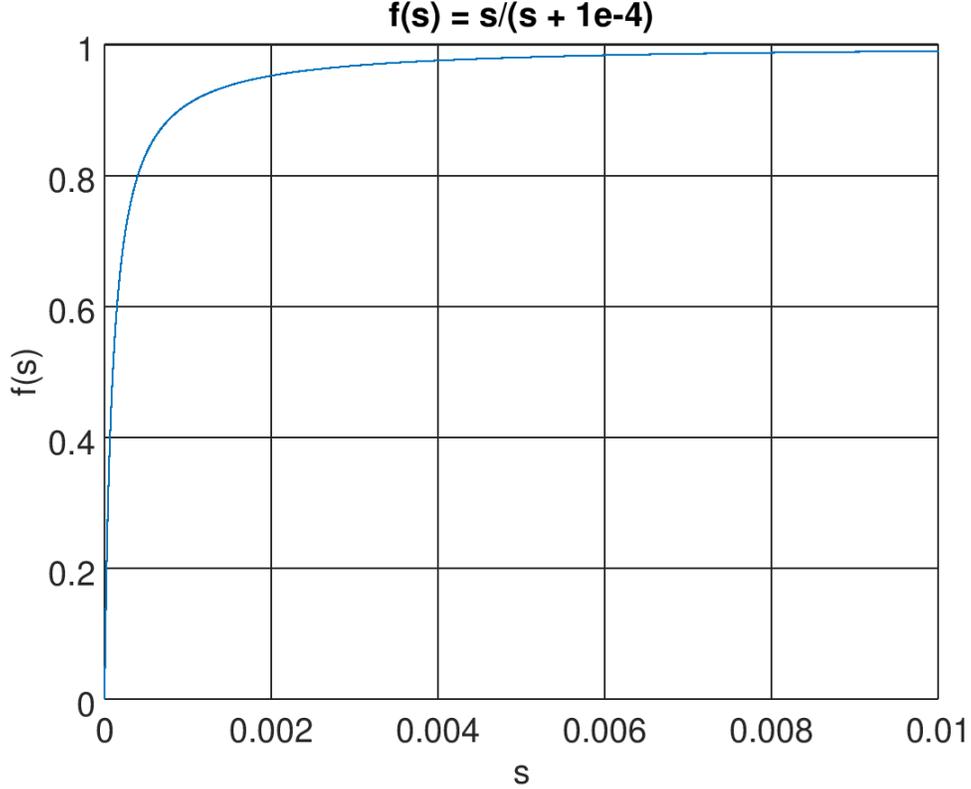}
 \caption{Schematic of correction term in PER method}
 \label{fig:PER}
\end{figure}

Besides the above error, the characteristic of the correction term can lead to a problem. As shown in Figure
\ref{fig:PER}, when $S$ approaches 0, the value and derivative of the function
$\left(\frac{s}{s+\epsilon}\right)$ change dramatically. The behavior of the function is similar to a
discontinuous function, which will lead to convergence failure in Newton iterations. Thus the pseudo K-values
are extremely sensitive to saturation, and very fine Newton step sizes on phase saturation are required. Also,
in Newton iterations, when $S$ is not at the order of $\epsilon$, the tangent slope is near zero. This causes
the saturation easily converging to a negative value, which is not acceptable. Therefore, a special treatment
of saturation in each Newton step is needed(\cite{oklany1992situ}):
\begin{gather}
S_o^{k+1} = \sqrt{\epsilon \cdot S_o^{k}}.\\
S_w^{k+1} = \sqrt{\epsilon \cdot S_w^{k}}.
\end{gather}
or
\begin{gather}
S_o^{k+1} = 0.5 S_o^{k}.\\
S_w^{k+1} = 0.5 S_w^{k}.
\end{gather}
Here, the superscript $k$ or $k+1$ denotes the k-th or (k+1)-th Newton step. If saturation is directly set to
a small value, a Newton iteration fails easily.

During the implementation, there is a modification to the PER method. \cite{crookston1979numerical} set a
model for only two components in oil phase. A pseudo K value is only applied to heavy oil component.
\cite{oklany1992situ} applied pseudo K values to all components in his simulator. In our simulator, pseudo K
values for the oil phase is only applied to one component, which is the heaviest component. There are two
reasons for this. First of all, during the physical process, lighter components will disappear first when the
oil phase is heated. This usually leaves the heaviest component to evaporate last. Molar fractions are
evaluated:
\begin{gather}
y_{LO} = K_{LO}(p, T) x_{LO},\label{eq.PER1}
\\
y_{HO} = \left(\frac{S_o}{S_o+\epsilon}\right)K_{HO}(p, T) x_{HO}.\label{eq.PER2}
\end{gather}
For a light component, its K value will increase earlier than a heavy component, and its molar fraction in the
oil phase $x$ will decrease earlier. After evaporation of the light component, the molar fraction in the oil
phase of the heavy component will approach 1. Thus a small value for $y$ of the light component results from a
small $x$ value. As for the heavy component, a small value for $y$ results from a small $S_o$ value. This
realization is close to the physical process.

On the other hand, this implementation deals with potential problems in numerical solution. When the oil phase
disappears ($S_o \longrightarrow 0$), the mass conservation equation of an oil component is changed to
$y^{n+1}=0$, where $n+1$ denotes the (n+1)-th time step. According to Equation \ref{eq.PER1} and
\ref{eq.PER2}, the equation system becomes
\begin{gather}
\begin{split}
x_{LO} &\simeq 0,\\
S_o &\simeq 0.\\
\end{split}
\end{gather}
However, if the correction term is added to each component, every single mass conservation equation for an oil component will be
\begin{gather}
S_o x_{O, i} \simeq 0.
\end{gather}
Since we know that $S_o \simeq 0$, the value of $x_{oil[i]}$ cannot be determined by the system. In the
solution process, it will result in linear dependence of the linear system as well as a singular Jacobian
system. An accurate solution to the linear system cannot be determined. Without this modified PER method, the
problem remains soluble when a restriction is applied to oil phase. When the oil phase almost disappears, both
convection term and reaction term should be set to zero, which prevent the oil phase decrease to zero. The oil
phase will remain an extremely small number, and the system will remain non-singular. However, this method
includes new discontinuity in the whole problem. The failures in Newton iteration are more common.  As a
result, this modified PER method in multicomponent system showed its effectiveness in solution process.

\subsubsection{Compressibility Factor of Real Gas}

Since an EOS equation is not introduced in our model for the calculation of phase equilibrium, a compressibility factor of gas is not acquired by this EOS approach. Another approach of this physical variable is implemented.

In our model, the Redlich-Kwong EOS (\cite{redlich1949thermodynamics}) is used to calculate the Z factor.
\begin{gather}
A = A(p, T) = 0.427480 \left(\frac{p}{p_{crit}}\right) \left(\frac{T_{crit}}{T}\right)^{2.5},\\
B = B(p, T) = 0.086640 \left(\frac{p}{p_{crit}}\right) \left(\frac{T_{crit}}{T}\right).
\end{gather}
IN addition, the following mixing method is applied:
\begin{gather}
a = \sum_{i} {y_i T_{crit,i} \sqrt{\frac{T_{crit,i}}{p_{crit,i}}}},\\
b = \sum_{i} {y_i \frac{T_{crit,i}}{p_{crit,i}}},\\
T_{crit} = \left(\frac{a^2}{b}\right)^{\frac{2}{3}},\\
p_{crit} = \frac{T_{crit}}{b}.
\end{gather}
Then, after we have the coefficients A and B, the compressibility factor of real gas satisfies the equation \begin{gather}
Z^3 - Z^2 + (A - B - B^2)Z - AB = 0.
\end{gather}
This equation is cubic. Therefore, there are three roots for the equation. Also, a root might be virtual. In this case, we choose the biggest real root. With the calculation of all the coefficients, the $Z$ factor is a function of $p$, $T$, $x_i$ and $y_i$:
\begin{gather}
Z = Z(p, T, x_i, y_i).
\end{gather}

\subsubsection{Density}

We introduce the molar density: the molar amount of certain material in unit volume in the simulator. This
usage ensures the unit of a mass conservation equation to be [mole/t]. The calculation here are consistent
with \cite{cmg2015starguide}.  For different phases, the calculation of densities varies a lot. We must
consider them separately.  First of all, for the gas phase, in 1834 Émile Clapeyron first stated the ideal gas
law. Since a volume taken by the gas phase is mainly affected by molecular collisions, the volume taken by a
certain amount of gas material is irrelevant to the type of the material itself. With the ideal gas law, under
a condition of certain pressure and temperature, the molar density of all kinds of gas remains the same.
(\cite{clapeyron1834memoir})

However, the ideal gas law is only an approximation to the real case. Though different EOS equations are all
approximations to the real case, they perform better than the ideal gas model when comparing to results from
the laboratory. For EOS equations, a compressibility factor $Z$ is introduced as a correction term for the
real gas model, while in the ideal gas model we have $Z = 1$. According to the discussion above, there is a
significant difference between the compressibility factor and $1$. Not only pressure and temperature also
different composition of the gas phase will lead to a significant difference in the $Z$ factor. For different
components, the critical pressure and critical temperature may differ, resulting in a significant difference
in gas behavior.

With a real gas equation considered, the density of the gas phase can be calculated as:
$$\rho_g = \rho_g(p, T, x_i, y_i) = \frac{p}{Z(p, T, x_i, y_i) \cdot R \cdot T}$$

The water phase only contains one water component in this model, so the calculation of the water density is simple:
\begin{gather}
\rho_w = \rho_w(p, T) = \rho_{w,ref} \exp (cp_w (p-p_{ref})-ct1_w (T-T_{ref}) \\
-\frac{ct2_w}{2}(T-T_{ref})^2 + cpt_w (p-p_{ref}) (T-T_{ref}))
\end{gather}
where $\rho_{w,ref}$ is the reference density of the water phase at the reference temperature and pressure.

For the oil phase, if only component $O[i]$ is in the oil phase, the density can be calculated similarly:
\begin{gather}
\rho_{O[i]} = \rho_{O[i]}(p, T) = \rho_{{O[i]},ref} \exp (cp_{O[i]} (p-p_{ref})-ct1_{O[i]} (T-T_{ref}) \\
-\frac{ct2_{O[i]}}{2}(T-T_{ref})^2 + cpt_{O[i]} (p-p_{ref}) (T-T_{ref}))
\end{gather}
However, under most circumstances, more than one component exist in the oil phase. The density of the oil phase should be considered with molar fractions of oil components. If there is one gmol of material in the oil phase, then for component $O[i]$, there will be $x_i$ gmol of it. Moreover, for one gmol of component $O[i]$ in the oil phase, it will take a volume of $\frac{1}{\rho_{O[i]}}$. Therefore, for unit gmol of material in the oil phase, it will take the volume of:
\begin{gather}
\frac{1}{\rho_o} = \sum^{n_{co}}_i \frac{1}{\rho_{O[i]}}x_i.
\end{gather}
Therefore, using a mixing principle, the harmonic average is introduced for the oil phase:
\begin{gather}
{\rho_o} = \rho_o(p, T, x_i) = 1 / \sum^{n_{co}}_i \frac{x_i}{\rho_{O[i]}(p, T)}.
\end{gather}
When the water phase contains multiple components, this equation is also valid for this phase.

\subsubsection{Viscosity}

The viscosity of heavy oil is very high, making it very hard to mobilize. With this characteristic, thermal recovery is introduced. The mechanism is that, when temperature is higher, the viscosity of heavy oil will decrease dramatically. For calculation, correlation methods are adopted here (\cite{cmg2015starguide}).

For the oil phase, if only component $O[i]$ is in the oil phase, it will have a viscosity:
\begin{gather}
\mu_{O[i]} = avisc_{O[i]} \exp{\left(\frac{bvisc_{O[i]}}{T}\right)}.
\end{gather}
This viscosity can be obtained from a viscosity table as well. As a mixture of different components, the oil phase viscosity is obtained by a logarithmic mixing rule:
\begin{gather}
\ln(\mu_o) = \sum_i^{n_{co}} x[i] \ln (\mu_{O[i]}(T)).
\end{gather}
Therefore, we see that
\begin{gather}
\mu_o = \mu_o (T, x_i) = \exp {\left(\sum_i^{n_{co}} x[i] \ln (\mu_{O[i]}(T))\right) } =\sum_i^{n_{co}} \left(\mu_{O[i]}(T)\right)^{x_i}.
\end{gather}

For the water phase, since it has only one component, the viscosity can be derived directly:
\begin{gather}
\mu_w = \mu_w (T) = avisc_w \exp{\left(\frac{bvisc_w}{T}\right)}.
\end{gather}

For the gas phase, however, the viscosity is usually much smaller than the oil viscosity or the water viscosity, and the effect of temperature on its viscosity is not as important as that for the oil phase or the water phase. Thus, another formula for the viscosity calculation is introduced:
\begin{gather}
\mu_{g,c} = \mu_{g,c}(T) = avg_c \cdot T ^ {bvg_c}.
\end{gather}
In this formula, the viscosity will not decline so fast with an increasing temperature. But under standard conditions, the viscosity of water is around 1 cp, while the viscosity of air is about 0.01 cp, much smaller than that of water. Therefore, the flow of gas will dominate.
According to a mixing rule, the molar mass of a component is included:
\begin{gather}
\mu_{g} = \mu_{g}(p, T, S_w, S_g, x_i, y_i) = \frac{\sum_c{\mu_{g,c} \cdot y_c\sqrt{M_c} }}{\sum_c{y_c\sqrt{M_c}}}.
\end{gather}

\subsubsection{Porosity}

Porosity is the ratio of the pore volume to the bulk volume in a porous medium, describing the volume
containing fluids. When pressure is high, due to the effort of fluids, pores are also enlarged. For a
non-isothermal model, the porosity is also influenced by temperature.
We define a coefficient as a total compressibility of porosity (\cite{chen2007reservoir}):
\begin{gather}
ctot = ctot(p, T) = cpor(p-p_{ref}) - ctpor(T-T_{ref}) + cptpor(p-p_{ref})(T-T_{ref}).
\end{gather}
This factor is a function of pressure and temperature. For the calculation of porosity, we have two approaches with this factor:

Linear:
\begin{gather}
\phi = \phi (p, T) = \phi_{ref} \cdot (1+ctot(p, T)).
\end{gather}

Nonlinear:
\begin{gather}
\phi = \phi (p, T) = \phi_{ref} \cdot e^{ctot(p, T)}.
\end{gather}
For both two approaches, porosity is a function of pressure and temperature.  Especially, for the ISC process,
we consider the effect of coke. In Coats’ simulator, when coke is generated, the solid phase appears. The
saturations of all four phases should be considered. (\cite{coats1980situ}) The phase constraint becomes
\begin{gather}
S_w+S_o+S_g+S_s = 1.
\end{gather}
In other words, when the solid phase appears, it will take up a part of the pore volume, but porosity will not be affected. Nevertheless, we will take the same treatment as in CMG's STAR. When the solid phase appears, it is immobile and thus takes a volume of the fluid space. In this case, coke has a similar effect as the matrix. This process can also be treated as porosity declining. With the formula
\begin{gather}
\phi_{f} = \phi - \phi_{coke},
\end{gather}
the porosity for fluids can be calculated by the original porosity and porosity is altered by coke. The concentration of coke is an unknown, which is defined by the amount of material (gmol) in unit volume ($m^3$). So the porosity taken by coke $\phi_{coke}$ can be calculated by
\begin{gather}
\phi_{coke} = C_c / \rho_{coke}
\end{gather}
In the solid phase, the volume can hardly be affected by pressure or temperature. Thus, the molar density of
coke can be taken as a constant. As a result, with the volume of coke being considered, porosity is a function
of pressure, temperature and coke concentration:
\begin{gather}
\phi = \phi(p, T, C_c)
\end{gather}
With this treatment of coke, our ISC simulator can also handle the process like in-situ coal gasification problems.

\subsubsection{Relative Permeabilities}

Just like any model containing three phases, relative permeabilities have the similar expression. Although a solid phase of coke is involved, this phase is not movable.
For the relative permeability of water $k_{rw}$, the relative permeability in a three-phase system equals the relative permeability in a water-oil system. Thus, from the SWT table (a water-oil relative permeability table), $k_{rw}$ can be obtained with interpolation. $k_{rw}$ is a function only of $S_{w}$:
\begin{gather}
k_{rw} = k_{rw}(S_w).
\end{gather}
In addition, the relative permeability of gas $k_{rg}$ in a three-phase system or an oil-gas system can be calculated in the same manner. Through interpolation with a SLT table (a liquid-gas relative permeability table), $k_{rg}$ can be calculated. Also, $k_{rg}$ is a function only of $S_{g}$:
\begin{gather}
k_{rg} = k_{rg}(S_g).
\end{gather}

As for the relative permeability of oil $k_{ro}$, there are several models available (\cite{corey1956three}; \cite{naar1961three}, 1961; \cite{stone1970probability}; \cite{delshad1989comparison}). In this simulator, we only include Stone’s model II (\cite{stone1973estimation}):
\begin{gather}
k_{ro} = k_{ro}(S_w, S_g)\\
= k_{rocw}\left[\left(\frac{k_{row}(S_w)}{k_{rocw}}+k_{rw}(S_w)\right)\left(\frac{k_{rog}(S_g)}{k_{rocw}}+k_{rg}(S_g)\right)-k_{rw}(S_w)-k_{rg}(S_g)\right].
\end{gather}
where $k_{rocw}$ is the oil-water two-phase relative permeability to oil at connate water saturation, $krog$ is the oil-gas two-phase relative permeability to oil, and $krow$ is the oil-water two-phase relative permeability to oil.
\begin{gather}
k_{rocw} = k_{row}(S_w = S_{wc}) = k_{rog}(S_g = 0).
\end{gather}
$krow$ and $krog$ are interpolated from a SWT table and a SLT table separately.

However, for our ISC simulator, temperature will change during the entire process. The temperature will affect the properties of non-isothermal flow such as relative permeabilities. During implementation, the Stone II model needs to be adjusted. However, due to the complexity of non-isothermal flow, the effect is not considered in our simulator at present. Also, in the upstream direction of a combustion front, the temperature is very high, which has a very big effect on relative permeabilities; while only the gas phase is present, the effect on total flow is not that important. Furthermore, in the downstream direction of a combustion zone, the temperature is not that high, and thus the effect of temperature is not necessarily involved.

\subsubsection{Non-isothermal Description}

Enthalpy is a measurement of energy in a thermodynamic system. It is the thermodynamic quantity equivalent to the total heat content of a system. It is equal to the internal energy of the system plus the product of pressure and volume.
For the gas phase, if only component $c$ exists, the gas enthalpy is calculated as follows (\cite{cmg2015starguide}):
\begin{gather}
H_{g,c} = H_{g,c}(T) = \int_{T_{ref}}^T {\left(cpg1_c + cpg2_c \cdot t + cpg3_c \cdot t^2 + cpg4_c \cdot t^3\right)}dt,
\end{gather}
$cpgi_c$, $i=1,2,3,4$, are constants for component $c$. Here, the component $c$ can refer to not only a gas component, but also an oil component or a water component. All the components must be able to exist in the gas phase, and the enthalpy of a component in the gas phase can generally be calculated by this formula.

Since the gas phase is a mixture, the gas enthalpy can be calculated by a weighted mean with gas molar fractions $y_c$:
\begin{gather}
H_g = H_g (p, T, S_w, S_g, x_i, y_i) = \sum_{c}^{N_c} {y_c H_{g,c}}.
\end{gather}

For the oil and water phases, the heat of vaporization should be considered. From the gas phase to a liquid phase, steam will condense, and part of the energy is released as heat. This part of the heat is measured by the heat of vaporization, which is the amount of energy (enthalpy) that must be added to a liquid substance to transform a unit amount of that substance into a gas. It can be calculated by:
\begin{gather}
H_{v,c} = H_{v,c} (T) = u(T_{crit,c} - T) \cdot hvr_c \cdot (T_{crit,c} - T)^{ev_c}.
\end{gather}
Here $u(x)$ denotes the Heaviside unit step function. The component’s enthalpy will decrease from the gas phase to a liquid phase when the temperature is lower than the critical temperature. This heat is also influenced by temperature.

Therefore, for a certain component $c$, its enthalpy in a liquid phase can be calculated as:
\begin{gather}
H_{c} = H_{c} (T) = H_{g,c} - H_{v,c}.
\end{gather}
where $H_{g,c}$ is the enthalpy of component $c$ in the gas phase, which can be calculated using the formula above.

As a result, for the water phase which only includes one component, the enthalpy is:
\begin{gather}
H_w = H_w (T) = H_{g,W} - H_{v,W}.
\end{gather}
For the oil phase, as a mixture, the enthalpy is:
\begin{gather}
H_o = H_o (p, T, x_i) = \sum_{i}^{n_{co}} {x_i (H_{g, O[i]} - H_{v, O[i]})}.
\end{gather}

Now, we have enthalpy for all three fluid phases, which are used to measure energy in heat convection.
Including the mechanical work, it is the energy carried by the material of a phase.

As heat accumulates, the internal energy must be used, because the pushing work is not necessarily included. For a liquid phase, its volume is set and cannot be changed much, leading to zero pushing work. Thus, the pushing work will only have a significant effect on the gas phase. Hence we have the formulas for three fluid phases (\cite{cmg2015starguide}):
\begin{gather}
U_w = U_w (T) = H_w - p/\rho_w,\\
U_o = U_o (p, T, x_i) = H_o - p/\rho_o,\\
U_g = U_g (p, T, S_w, S_g, x_i, y_i) = H_g - p/\rho_g.
\end{gather}

For our simulator, since the solid phase and rock have no mobility, the related properties are not included,
such as density. Both have internal energy, and for heat conduction, this part of the energy will also play an
important role in the whole process.

For rock, a similar formula is used:
\begin{gather}
U_r = U_r (T) = cp1_r (T-T_{ref}) + \frac{cp2_r}{2}(T^2 - T_{ref}^2).
\end{gather}
One thing to notice is that the internal energy for rock has a unit of energy per unit volume, while others have energy per unit amount of material.

As for the solid phase, if coke has a heat capacity, which is the only component in the solid phase, then the enthalpy can be calculated as:
\begin{gather}
U_c = U_c(T) = cp_c (T-T_{ref}).
\end{gather}

\subsubsection{Well Modeling}

A Peaceman's model is adopted for well modeling. A well may have many perforations. For each perforation at a grid cell, its well rate for phase $\alpha$, $Q_{\alpha} = Vq_{\alpha}$, is calculated by the following formula (\cite{peaceman1978interpretation}):
\begin{gather}
Q_{\alpha, well} = WI\frac{\rho_{\alpha}k_{r\alpha}}{\mu_{\alpha}}\left(bhp - p_\alpha -\gamma_{\alpha}g(z_{bh} - z)\right),
\end{gather}
where WI is the well index. For a vertical well, it can be defined as:
\begin{gather}
WI= \frac{2\pi h_3 \sqrt{k_{11}k_{22}}}{\ln(\frac{r_e}{r_w})+s}.
\end{gather}
A well index defines the relationship among a well bottom hole pressure, a flow rate and a grid block pressure. $bhp$ is the bottom hole pressure defined at the reference depth z, $z_bh$ is the depth of the perforation in grid cell, and $p_\alpha$ is the phase pressure in grid block $m$.

From the mass conservation equation of component $c$, the well term can be obtained:
\begin{gather}
Q_{c, well} = Vq_{c, well} = \sum_{\alpha}^{N_\alpha}Q_{\alpha, well}\cdot x_{c,\alpha}.
\end{gather}
From the energy conservation equation, the well term can be obtained:
\begin{gather}
Q_{well} = \sum_{\alpha}^{N_\alpha}Q_{\alpha, well}\cdot H_{\alpha}.
\end{gather}

If a well is added to the reservoir model, the bottom hole pressure of a well becomes an unknown. Due to various operation constraints,  a fixed bottom hole pressure, a fixed oil rate, a fixed water rate or a fixed liquid rate may be applied to a well at different time stages. When the fixed bottom hole pressure condition is applied to a well, the well equation is
\begin{gather}
bhp = c.
\end{gather}
Here $c$ is a constant pressure.

The fixed rate operation for phase $\alpha$ is described by the following equation:
\begin{gather}
\sum_{perf} \left( Q_{\alpha, well} \right)_{perf} = c,
\end{gather}
where $c$ is a constant rate and known.

The total fixed rate operation is described by the following equation:
\begin{gather}
\sum_{perf} \left( Q_{w, well} + Q_{o, well} + Q_{g, well}  \right)_{perf} = c.
\end{gather}

Different from other equations, the well equations are handled by the last processor when running a parallel
simulation. As discussed before, a given grid will be divided into different subdomains, stored on different
nodes. All the cells and related properties are stored together. Thus, for one single processor, all the
equations in cells in this part are handled by this processor.  In one scenario, cells with perforations of a
certain well should not be divided into different parts. Otherwise, for a certain processor, when assembling
well equations, communications between different processors are required, which will increase the overhead of
the total assembling time and solution time. However, the loading balance solution we use makes it kind of
difficult to consider the integrity of perforation cells of one well. On the other side, comparing both
effectiveness and quality of a loading balance method, the overhead resulting from well perforation cells is
acceptable.  Therefore, to have the best performance and quality of domain partitioning, wells are not
considered in separate processors. All the well equations will be handled only by the last processor. With
message passing from/to other processors, the well equations can be assembled.

\section{Numerical Methods}
\label{sec-num}

The partial differential equations and constraints are discretized based on structured or unstructured grids.
The main purposes of a numerical discretization method include providing geometrical information and computing
residuals and derivatives on each element or control volume.  Detailed information will be introduced in the
following sections. 

\subsection{Discretization Scheme}

Typically, different time discretization (time marching) methods can be applied. Explicit (or forward
difference, forward Euler) and implicit (or backward difference, backward Euler) schemes are the two most
popular methods and are both first-order accurate in time. The Crank-Nicolson scheme is second-order accurate.
An explicit scheme is the simplest method in which all fluxes and sources are evaluated by using their
previous time step values. It is stable provided that a time step is sufficiently small (conditionally
stable). Therefore, if a time step is large, a stable constraint must be determined to restrict and modify the
time step. Under a given set of conditions, a time step cannot exceed this constraint, which will prevent the
time step from increasing and the total number of iterations from deceasing. Therefore, for an explicit
scheme, its performance may not be very good with a large number of time iterations.
For an implicit scheme, its fluxes and sources are evaluated in terms of the unknown variable values at the
new time step. So both the current state of a system and the later one are involved. Although an implicit
scheme requires more memory (a linear system of algebraic equations needs to be stored and solved), the use of
the implicit Euler scheme allows for large time steps to be taken, increasing the overall performance of a
simulator. However, a time step may be limited due to the severely nonlinear nature of reservoir dynamics
created by multi-phase, multi-component mass and heat flow with chemical reactions through heterogeneous
porous media.

As a result, the time discretization is carried out with an implicit scheme in our simulator. As for spatial
discretization, typically, the Finite Difference Method (FDM), Finite Element Method (FEM), and Finite Volume
Method (FVM) can be applied in the ISC process. In our approach, we adopt the FDM due to its good performance
with simplicity of parallelization and development.

Before discretization, let us illustrate some concepts and quantities. First, we have the following relationship in time steps:
$$t^{n+1} = t^{n} + \Delta t^n.$$
Here, superscript $n$ denotes the previous time step and $n + 1$ the current time.

Second, recalling the mass conservation equations (\ref{eq4.2}) and energy conservation equation (\ref{eq4.3}), we introduce the potential for each fluid phase, defined as
\begin{gather*}
\Phi_{w} = p_{w} - \gamma_{w} z\\
\Phi_{o} = p_{o} - \gamma_{o} z\\
\Phi_{g} = p_{g} - \gamma_{g} z
\end{gather*}
Therefore, Darcy’s flow equations can be rewritten in the form of
\begin{gather*}
\vec{u}_w = - k \frac{k_{rw}}{\mu_w} \nabla \Phi_{w},\\
\vec{u}_o = - k \frac{k_{ro}}{\mu_o} \nabla \Phi_{o},\\
\vec{u}_g = - k \frac{k_{rg}}{\mu_g} \nabla \Phi_{g}.
\end{gather*}

Third, we introduce the difference sign $\Delta$ and take a potential as an example: $$\Delta \Phi_{\alpha,
face} = (\Phi_{\alpha})_c - (\Phi_{\alpha})_{face}.$$ Here, a face is a direction in space. Typically, it is
defined by a Cartesian coordinate system.

Along with the $x$, $y$, and $z$ directions and three opposite directions, we have six directions defined.
Each direction is denoted by an integer from 0 to 5. A face can denote the meaning of front, back, left,
right, up or down. $c$ here denotes a center cell. Thus, a face here actually refers to the neighboring cell
in the direction of the ``face''.  Then we have the discretization of the governing equations. Here, we omit
the superscript $n + 1$, and $n$ is preserved. Also, if not noted by a “face”, all properties and equations
are regarding the center cell $c$. Otherwise, they are regarding the neighboring cell in the “face” direction.

First, we obtain mass conservation equations in new forms. For the water component $W$:
\begin{gather*}
F_W = - \sum_{face = 0}^{5} \left(\frac{Ak}{h}\right)_{avg}\left(\frac{k_{rw}\rho_w}{\mu_w}\right)_{avg} \Delta \Phi_{w, face}\\
+\frac{V}{\Delta t} \left[\phi \rho_w S_w - (\phi \rho_w S_w)^n \right]\\
- \sum_{face = 0}^{5} \left(\frac{Ak}{h}\right)_{avg}\left(\frac{k_{rg}\rho_g}{\mu_g}y_W\right)_{avg} \Delta \Phi_{g, face}\\
+\frac{V}{\Delta t} \left[\phi \rho_g S_g y_W - (\phi \rho_g S_g y_W)^n \right]\\
- Vq_{W,well} - Vq_{W,reac}.
\end{gather*}
Here, $V$ denotes the volume of a center cell, $A$ denotes a cross-sectional area, and $h$ denotes the length of an edge of the cell. The water phase and the gas phase have mass transfer so these two phases need to be involved in this equation.

Similarly, for the oil component $O[i]$:
\begin{gather*}
F_{O[i]} = - \sum_{face = 0}^{5} \left(\frac{Ak}{h}\right)_{avg}\left(\frac{k_{ro}\rho_o}{\mu_o}x_i\right)_{avg} \Delta \Phi_{o, face}\\
+\frac{V}{\Delta t} \left[\phi \rho_o S_o x_i - (\phi \rho_o S_o x_i)^n \right]\\
- \sum_{face = 0}^{5} \left(\frac{Ak}{h}\right)_{avg}\left(\frac{k_{rg}\rho_g}{\mu_g}y_{O[i]}\right)_{avg} \Delta \Phi_{g, face}\\
+\frac{V}{\Delta t} \left[\phi \rho_g S_g y_{O[i]} - (\phi \rho_g S_g y_{O[i]})^n \right]\\
- Vq_{O[i],well} - Vq_{O[i],reac}.
\end{gather*}

But for the gas component $G[i]$, it only exists in the gas phase so that
\begin{gather*}
F_{G[i]} = - \sum_{face = 0}^{5} \left(\frac{Ak}{h}\right)_{avg}\left(\frac{k_{rg}\rho_g}{\mu_g}y_{G[i]}\right)_{avg} \Delta \Phi_{g, face}\\
+\frac{V}{\Delta t} \left[\phi \rho_g S_g y_{G[i]} - (\phi \rho_g S_g y_{G[i]})^n \right]\\
- Vq_{G[i],well} - Vq_{G[i],reac}.
\end{gather*}

For the coke equation, we have:
\begin{gather*}
F_{Coke} = \frac{V}{\Delta t} \left[C_c - (C_c)^n \right] - Vq_{c,reac}.
\end{gather*}

Similarly, for the energy conservation equation, we have the discrete form:
\begin{gather*}
F_{E} =
- \sum_{face = 0}^{5} \left(\frac{Ak}{h}\right)_{avg}\left(\frac{k_{rw}\rho_w}{\mu_w}H_w\right)_{avg} \Delta \Phi_{w, face}\\
- \sum_{face = 0}^{5} \left(\frac{Ak}{h}\right)_{avg}\left(\frac{k_{ro}\rho_o}{\mu_o}H_o\right)_{avg} \Delta \Phi_{o, face}
- \sum_{face = 0}^{5} \left(\frac{Ak}{h}\right)_{avg}\left(\frac{k_{rg}\rho_g}{\mu_g}H_g\right)_{avg} \Delta \Phi_{g, face}\\
+ \frac{V}{\Delta t} \left[\phi \rho_w S_w U_{w} - (\phi \rho_w S_w U_{w})^n \right]
+ \frac{V}{\Delta t} \left[\phi \rho_o S_o U_{o} - (\phi \rho_o S_o U_{o})^n \right]
+ \frac{V}{\Delta t} \left[\phi \rho_g S_g U_{g} - (\phi \rho_g S_g U_{g})^n \right]\\
+ \frac{V}{\Delta t} \left[(1-\phi) U_{r} - ((1-\phi) U_{r})^n \right]
+ \frac{V}{\Delta t} \left[C_c U_{c} - (C_c U_{c})^n \right]\\
- \sum_{face = 0}^{5} \left(\frac{A}{h}\right)_{avg}\left(K_T\right)_{avg} \Delta T_{face}\\
- VQ_{well} - VQ_{reac} + VQ_{loss}.
\end{gather*}

For the gas and oil composition constraints, we also have the discrete form:
\begin{gather*}
F_{GS} = \sum_{c}y_c - 1,\\
F_{OS} = \sum_{i}^{nco}x_i - 1.
\end{gather*}
Therefore, all the discretization is finished for all cell equations.

The properties of a grid, rock, and fluids, such as thickness, permeability, and porosity, can significantly
differ from one block to another. The properties are often given only at the centers of cells, but the
transmissibilities are computed at the boundaries of cells. For transmissibilities at grid block boundaries,
proper average techniques between two adjacent blocks should be taken into consideration.
For transmissibility in a ``face'' direction, a different method of calculating averages should be applied. For the first term, grid and rock properties are depicted, and a harmonic average should be used:
$$\left(\frac{Ak}{h}\right)_{avg} = \frac{2\left(\frac{Ak}{h}\right)_{c}\left(\frac{Ak}{h}\right)_{face}}{\left(\frac{Ak}{h}\right)_{c} + \left(\frac{Ak}{h}\right)_{face}}$$
The fluid properties are described in the second term in transmissibility. Especially for multi-phase flow, upstream weighting should be applied to calculate rock/fluid properties. This prevents a physically incorrect solution. With a fluid potential defined,
\begin{gather*}
\left(\frac{k_{r\alpha}\rho_\alpha}{\mu_\alpha}\right)_{avg} =
\left\{
    \begin{array}{lr}
    \left(\frac{k_{r\alpha}\rho_\alpha}{\mu_\alpha}\right)_{c}, & \mbox{if} (\Phi_{\alpha})_c > (\Phi_{\alpha})_{face}  \\
    \left(\frac{k_{r\alpha}\rho_\alpha}{\mu_\alpha}\right)_{face}, & \mbox{if} (\Phi_{\alpha})_c < (\Phi_{\alpha})_{face}
    \end{array}
\right.
\end{gather*}
If $(\Phi_{\alpha})_c > (\Phi_{\alpha})_{face}$, we call cell $c$ the upstream cell, and the cell face is the
downstream block. We should take the properties of the upstream cell to be the term at the boundary
(\cite{ertekin2001basic}; \cite{chen2007reservoir}).

In the in-situ combustion process, the gas phase will not disappear for non-condensable gas injected and
produced during chemical reactions. Such gases include oxygen, carbon oxides, and nitrogen. A saturated state
for a thermal model always exists. With the correction terms of K-values, the liquid phases also exist.
Therefore, the structure of this system does not change during calculations.

\subsection{Decoupling Method}

Before solving the linear system from Newton method, a decoupling method is applied to the system, 
in which the decoupling method
can accelerate the solution.
In general, the linear systems from the ISC model are hard to solve, in which a big condition number results 
from two aspects: coupling in a cell matrix and a large scale of grid cells.

First, for the matrix in a single cell, when the number of components is large, the coupling phenomenon can be
extremely obvious, making its condition number extremely big. If the condition number gets bigger, it
indicates that the matrix is closer to a singular matrix, which makes a linear system very hard to solve.
Second, for the adoption of pseudo K-values as discussed in the sub-section of K-values, the derivatives of
these pseudo K-values with respect to saturation can be extremely large when the saturation of water or oil is
very small. As shown in Figure \ref{fig:PER}, the tangent at $s=0$ is nearly parallel to the $y$-axis. For a
correlation term, if we take function the 
$$f(s) = \frac{s}{s+\epsilon},$$
then
$$f'(s) = \frac{\epsilon}{(\epsilon + s)^2}.$$ 
Therefore, $$f'(0) = \frac{1}{\epsilon}, $$ indicating that when the oil
phase or water phase is about to disappear, the entries in the Jacobian matrices $\frac{\partial F_i}{\partial
S_w}$ and $\frac{\partial F_i}{\partial S_g}$ can be extremely big compared to other entries.
This phenomenon
also makes a matrix ill-conditioned. In the implementation, we set the $crk\_mod$ to be not too small.
According to the literature and our tests, a number around 1e-4 would lead to a good balance between the
condition number and relative error of mass conservation. However, with an application of the PER method, the
matrix is still not well-conditioned. In this case, a preconditioner is required.

Third, the primary variable selection and variable alignment are considered in many reservoir simulators with
the aim to make the nature of a matrix better. One example is to make a matrix diagonally dominant or positive
definite. But in a process of ISC, changes among phases are violent. The correspondence between equations and
unknowns set at the beginning may preserve the good nature of a matrix, but it may not be suitable for a new
state of equations as time goes. Therefore, a good arrangement may be damaged. Under this circumstance, the
matrix may not be so well-conditioned.
Many approximate
decoupling methods are proposed, such as the quasi-IMPES method, classical IMPES 
method\cite{gries2015system,jiang2008techniques,lacroix2000iterative},
the ABF (Alternate Block Factorization) method \cite{bank1989alternate}
and local QR decomposition for
decoupling\cite{klie1996two,lacroix2000iterative}.

The coupling problem is caused not only by the pressure-saturation coupling but also caused by
multi-components. Therefore, a general inter-equation couplings exists in a system, which ABF can handle.
The ABF method is achieved by scaling a Jacobian matrix $J$ such that its diagonal blocks become the unit
block. For the linear system $J\delta\vec{u} = -\vec{f}$, $J=(J_{i,j})_{ncell \times ncell}$, where $J_{i,j}$
is the main cell matrix of the first cell. Then we define
\begin{gather*}
C_L =
\left[
    \begin{array}{ccc}
    (J_{1,1})^{-1}&  &  \\
     & \ddots &  \\
     &  & (J_{ncell,ncell})^{-1}
    \end{array}
\right].
\end{gather*}
A new linear system can be obtained: $$C_LJ\delta\vec{u} = -C_L\vec{f}.$$ With this decoupling, the diagonal
blocks of this Jacobian matrix will change to the unit block.

When implementing the ABF decoupling method, the Gauss-Jordan transformation is employed. Different form
multiplied by an inverse matrix directly, each diagonal block in the Jacobian matrix is transformed in less
steps than traditional ABF method. The complexity of computations can be reduced from $ncell \cdot O(nequ^3)$
to $ncell \cdot O(nequ^2)$.  In the process of Gauss-Jordan transformation, three types of elementary row
operations are employed to achieve this goal. The main difference between this method and ABF is that
multiplications between blocks are avoided. This method is called the Gauss-Jordan elimination through
pivoting.

\section{Numerical Experiments}
\label{sec-results}

An in-house platform has been implemented by our group
\cite{liu2017dynamic, liu2016family, liu2016performance, liu2015parallel}.
The platform provides parallel computing, gridding, load balancing, distributed-memory matrix and vector,
linear solver and a family of CPR-like preconditioners\cite{liu2016family}, parallel input and output through
MPI-IO, visualization, keyword parsing and well modeling. Several reservoir simulators have been developed
based on thi platform, such as black oil, compositional, thermal and the ISC simulator proposed in this paper.
In this section, some results of different experiments will be presented and some discussions will be given.
Also, validations will be shown to examine the legitimacy of the numerical model. Mainly, results from our
simulator RSGISC will be compared with CMG STARS.

\subsection{Combustion Tube Experiment}

First of all, a combustion tube experiment is presented as a 1D case. All the data was constructed from CMG
case stdrm001.dat.
In this case, the validity is shown first by the physical properties profile and second by the comparison
between the simulator and CMG STARS.

\subsubsection{Model Description}

The reservoir is 0.1602ft long, 0.1602ft wide, and 2.6458ft high, and performed as a one-dimensional model, which is partitioned into 12 cells vertically. Air is injected from the top into the tube and fluids are produced at the bottom. Oil sample in this combustion tube test is light oil sample (24.26 ${}^\circ API $ at $^\circ F$). The viscosity is 22.6 $cp$, which is not very high. Therefore, this sample is suitable for Crookston's model, where the combustion is mainly for light oil. LO and HO denote the light oil component and heavy oil component. IR denotes the inert gas pseudo-component, where nitrogen and carbon oxides exist.
The set of data is listed below:

\begin{table}[!htb]
\centering
\begin{tabular}{l c}
\hline
\textbf{Initial condition} &  \\
\hline
$k_{x,y,z}\ (md)$  & 12700, 12700, 12700 \\
$\phi$ & 0.4142 \\
$p\ (psi)$ & 2014.7 \\
$T\ (^\circ F)$ & 100 \\
$S_{w, o, g}$ & 0.178, 0.654, 0.168 \\
$x_{LO, HO}$ & 0.744 0.256 \\
$y_{W, LO, HO, O2, IR}$ & 0.0, 0.0, 0.0, 0.21, 0.79\\
\hline
\end{tabular}
\caption{Input data for tube model.}
\label{Tab:input11}
\end{table}

\begin{table}[!htb]
\centering
\begin{tabular}{l c c c c c c}
\hline
\textbf{Properties} & \textbf{H2O} & \textbf{LO} & \textbf{HO} & \textbf{O2} & \textbf{IR} & \textbf{Coke} \\
$M\ (lb/lbmole)$ & 18 & 156.7 & 675 & 32 & 40.8 & 13 \\
$p_{crit}\ (psi) $ & 3155 & 305.7 & 120 & 730 & 500\\
$T_{crit}\ ({}^\circ F)$ & 705.7 & 651.7 & 1138 & -181 & -232\\
\hline

$\rho_{ref} (lbmole/ft^3)$ & 3.466 & 0.3195 & 0.0914 & & &57.2 \\
$cp\ (1/psi)$ & 3e-6 & 5e-6 & 5e-6 & & & \\
$ct1\ (1/{^\circ F})$ & 1.2e-4 & 2.839e-4 & 1.496e-4 & & & 4.06\\
\hline

$cpg1\ (Btu/({^\circ F} \cdot lbmol))$ & 7.613 & -1.89 & -8.14 & 6.713 & 7.44\\
$cpg2\ (Btu/({^\circ F}^2 \cdot lbmol))$ & 8.616e-4 & 0.1275 & 0.549 & -4.883e-7 & -0.0018\\
$cpg3\ (Btu/({^\circ F}^3 \cdot lbmol))$ & 0 & -3.9e-5 & -1.68e-4 & 1.287e-6 & 1.975e-6\\
$cpg4\ (Btu/({^\circ F}^4 \cdot lbmol))$ & 0 & 4.6e-9 & 1.98e-8 & -4.36e-10 & -4.78e-10\\
$hvr\ (Btu/({^\circ F}^{ev} \cdot lbmol))$ & 1657 & 1917 & 12198 & & & \\
$ev$ & 0.38 & 0.38 & 0.38  & & & \\
\hline
$avg\ (cp/{^\circ F})$ & 8.822e-6 & 2.166e-6 & 3.926e-6 & 2.196e-4 & 2.127e-4\\
$bvg$ & 1.116 & 0.943 & 1.102 & 0.721 & 0.702\\
$avisc\ (cp)$ & 4.7352e-3 & 4.02e-4 & 4.02e-4 & & & \\
$bvisc\ ({^\circ F})$ & 2728.2 & 6121.6 & 6121.6 & & & \\
\hline
$kv1\ (psi)$ & 1.7202e6 & 1.4546e5 & 2.7454e5 & \multicolumn{3}{c}{\mbox{$kv3$ and $kv4$ are all 0}}\\
$kv4\ ({^\circ F})$ & -6869.59 & -4458.73 & -8424.83 & & & \\
$kv5\ ({^\circ F})$ & -376.64 & -387.78 & -205.69 & & & \\
\hline
\end{tabular}
\caption{Input data for tube model (cont'd).}
\label{Tab:input3}
\end{table}

\begin{table}[!htb]
\centering
\begin{tabular*}{0.8\textwidth}{l c}
\hline
\textbf{Reaction data} & \\\hline
$LO + 14.06 O_2 \longrightarrow 11.96 IR + 6.58 H_2 O$ & $r_1$ \\
$HO + 60.55 O_2 \longrightarrow 51.53 IR + 28.34 H_2 O$ & $r_2$ \\
$HO \longrightarrow 2.154 LO + 25.96 Coke$ & $r_3$ \\
$Coke + 1.18 O_2 \longrightarrow 1 IR + 0.55 H_2 O$ & $r_4$\\
$A_R\ (1/(psi \cdot day))$ & 7.248e11, 7.248e11, 1.00008e7, 1.00008e4\\
$E_{aR}\ (Btu/lbmol)$ & 59450, 59450, 27000, 25200\\
$H_R\ (Btu/lbmol)$ & 2.9075e6, 1.2525e7, 4.0e4, 2.25e5\\
\end{tabular*}
\\
\begin{tabular*}{0.8\textwidth}{c c c}
\hline
\textbf{Well conditions} & & \\\hline
Injector & air rate (O2:N2 = 0.21:0.79) $(ft^3/hr)$ & 0.554  \\
 & wi $(ft \cdot md)$ & 5.54 \\
 & pinjw $(psi)$ & 10000.0 \\
 & tinjw $(^\circ F)$ & 70 \\
 & heat rate $(Btu/day)$ & 4800 \\
 & heat stop time $(day)$ & 0.02083333333 \\
Producer & bhp $(psi)$ & 2014.7 \\
\hline
\end{tabular*}
\caption{Input data for tube model (cont'd).}
\label{Tab:input6}
\end{table}

\subsubsection{Validation with CMG STARS}

Before adopting a simulator for practical purposes, the establishment of confidence in its predictive
capability is important. To achieve this goal, its simulation results are compared with those by the 
CMG STARS(\cite{cmgstars}; \cite{cmg2015starguide}).

\begin{figure}[!htb]
 \centering\includegraphics[width=0.75\textwidth]{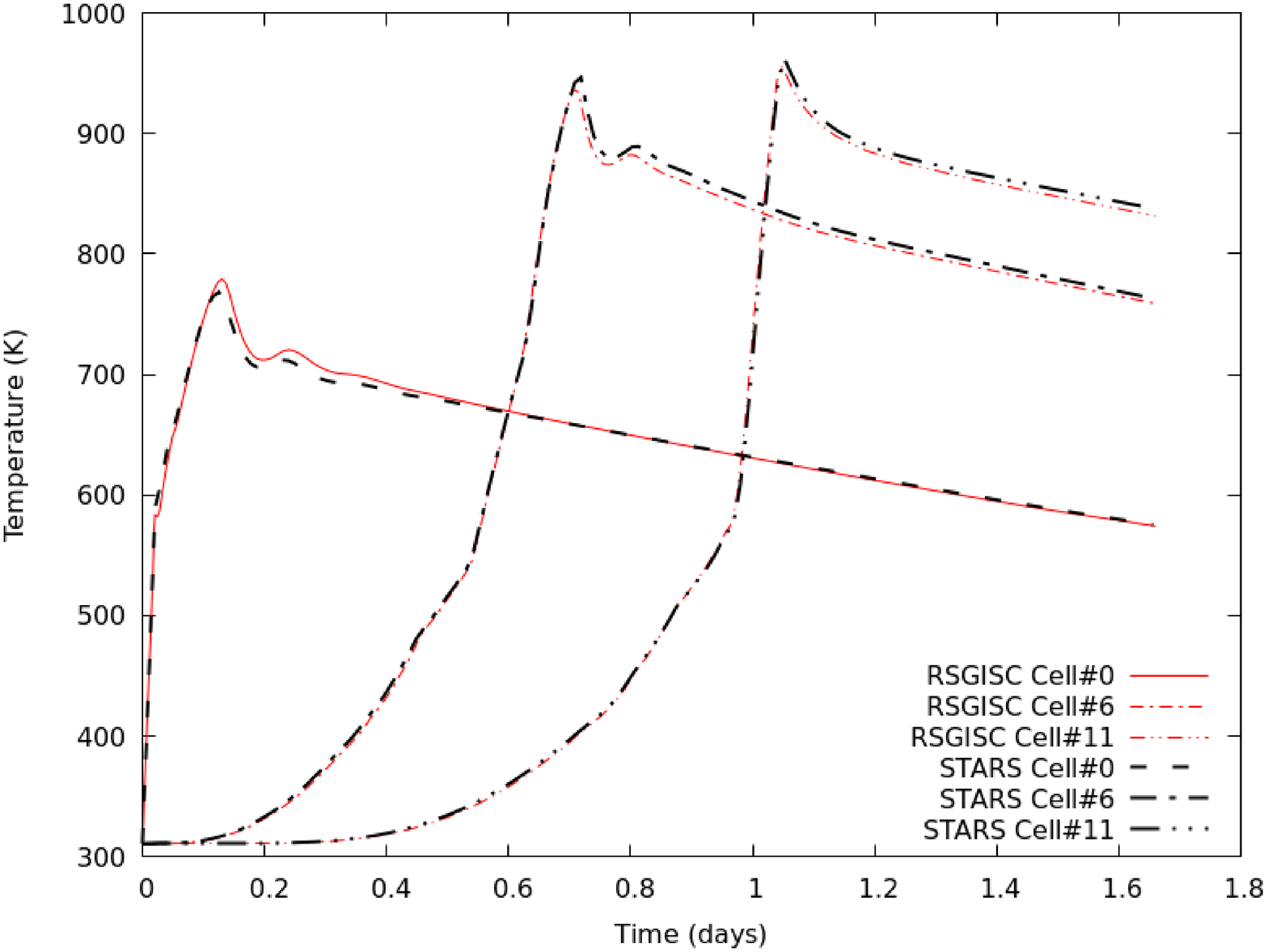}
 \caption{Combustion Tube Experiment: Temperature profile with results from CMG STARS.}
 \label{fig:ex1_cmg_t}
\end{figure}

\begin{figure}[!htb]
 \centering\includegraphics[width=0.75\textwidth]{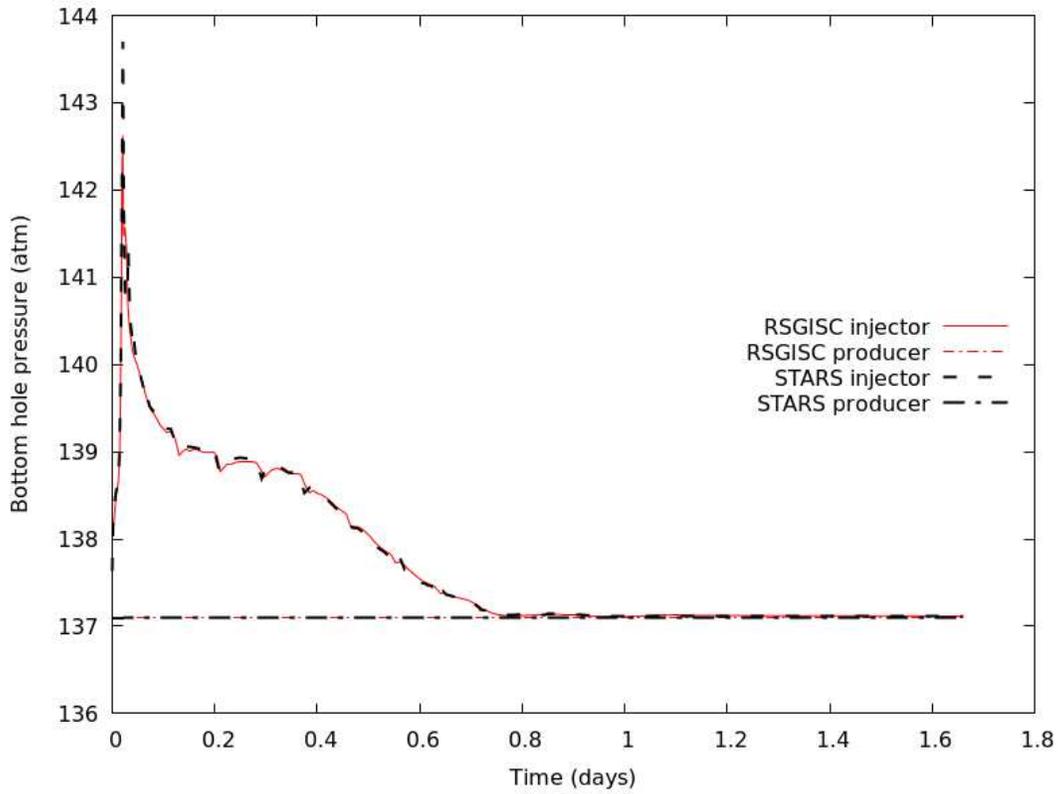}
 \caption{Combustion Tube Experiment: Bottom hole pressure profile with results from CMG STARS.}
 \label{fig:ex1_cmg_bhp}
\end{figure}

\begin{figure}[!htb]
 \centering\includegraphics[width=0.75\textwidth]{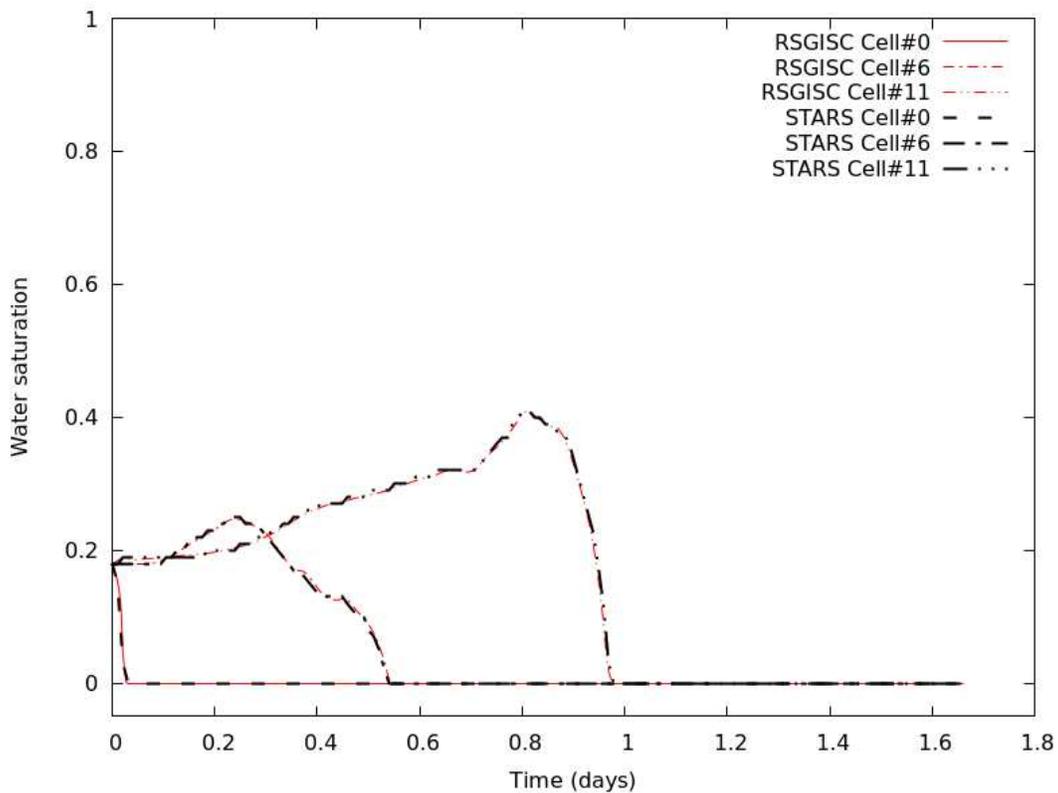}
 \caption{Combustion Tube Experiment: Water saturation profile with results from CMG STARS.}
 \label{fig:ex1_cmg_sw}
\end{figure}

\begin{figure}[!htb]
 \centering\includegraphics[width=0.75\textwidth]{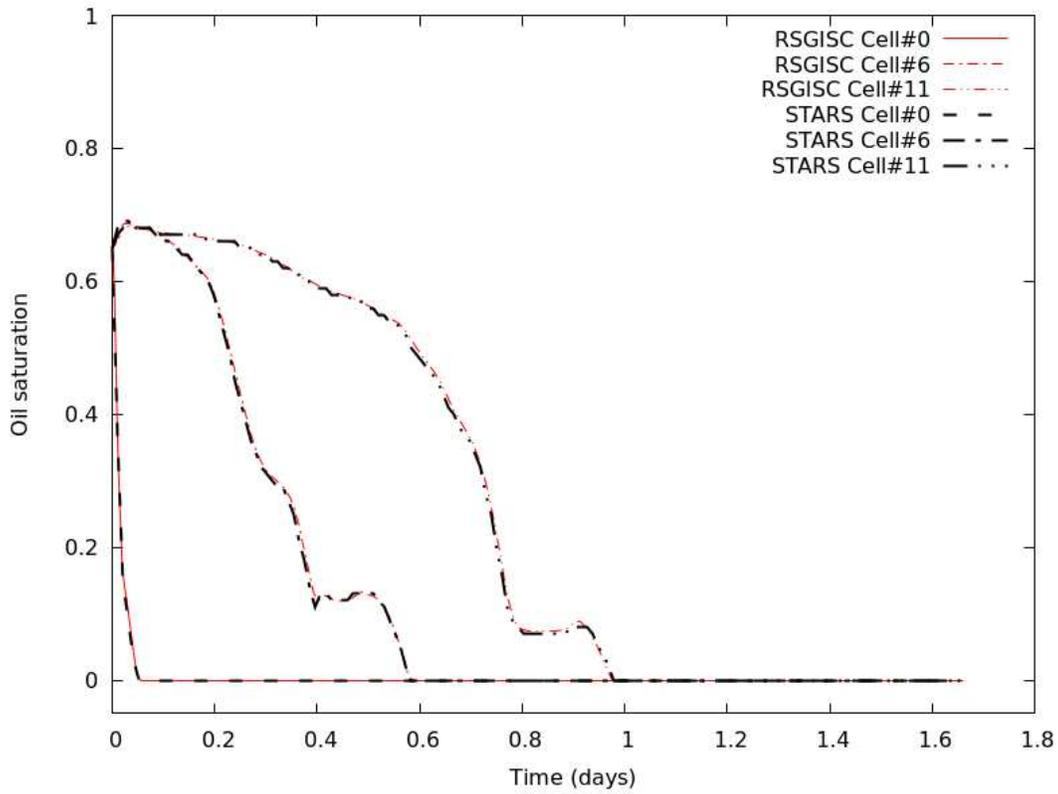}
 \caption{Combustion Tube Experiment: Oil saturation profile with results from CMG STARS.}
 \label{fig:ex1_cmg_so}
\end{figure}

\begin{figure}[!htb]
 \centering\includegraphics[width=0.75\textwidth]{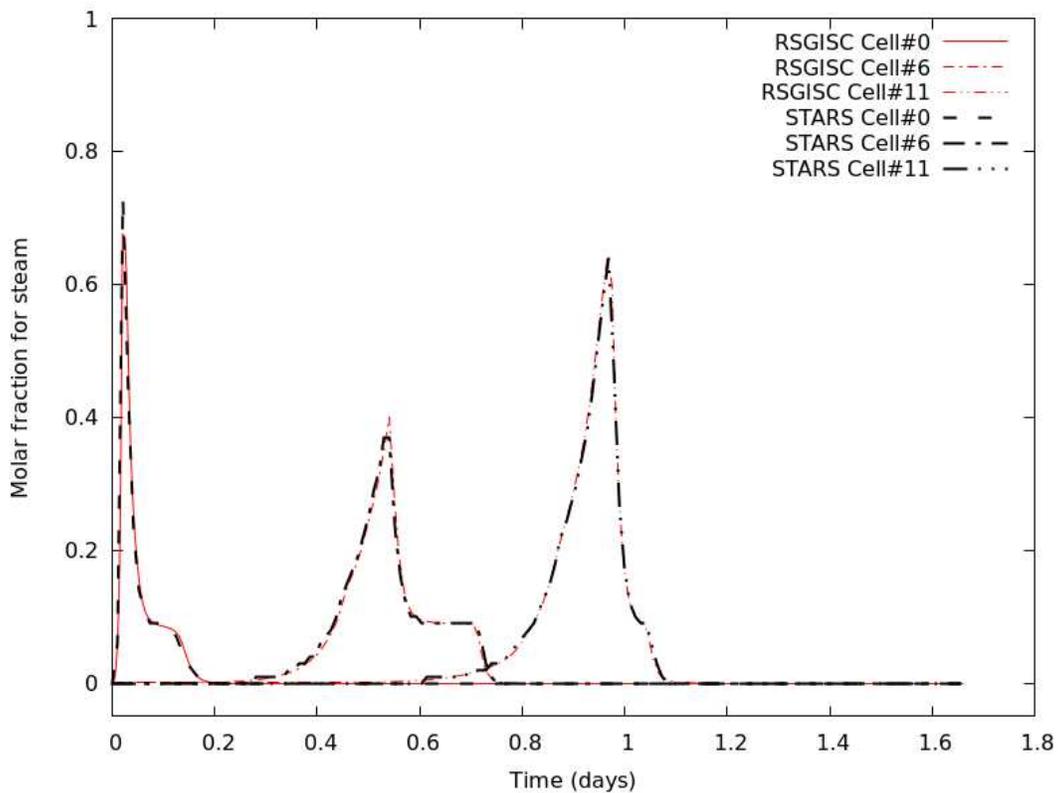}
 \caption{Combustion Tube Experiment: Gas molar fraction of steam profile with results from CMG STARS.}
 \label{fig:ex1_cmg_y0}
\end{figure}

\begin{figure}[!htb]
 \centering\includegraphics[width=0.75\textwidth]{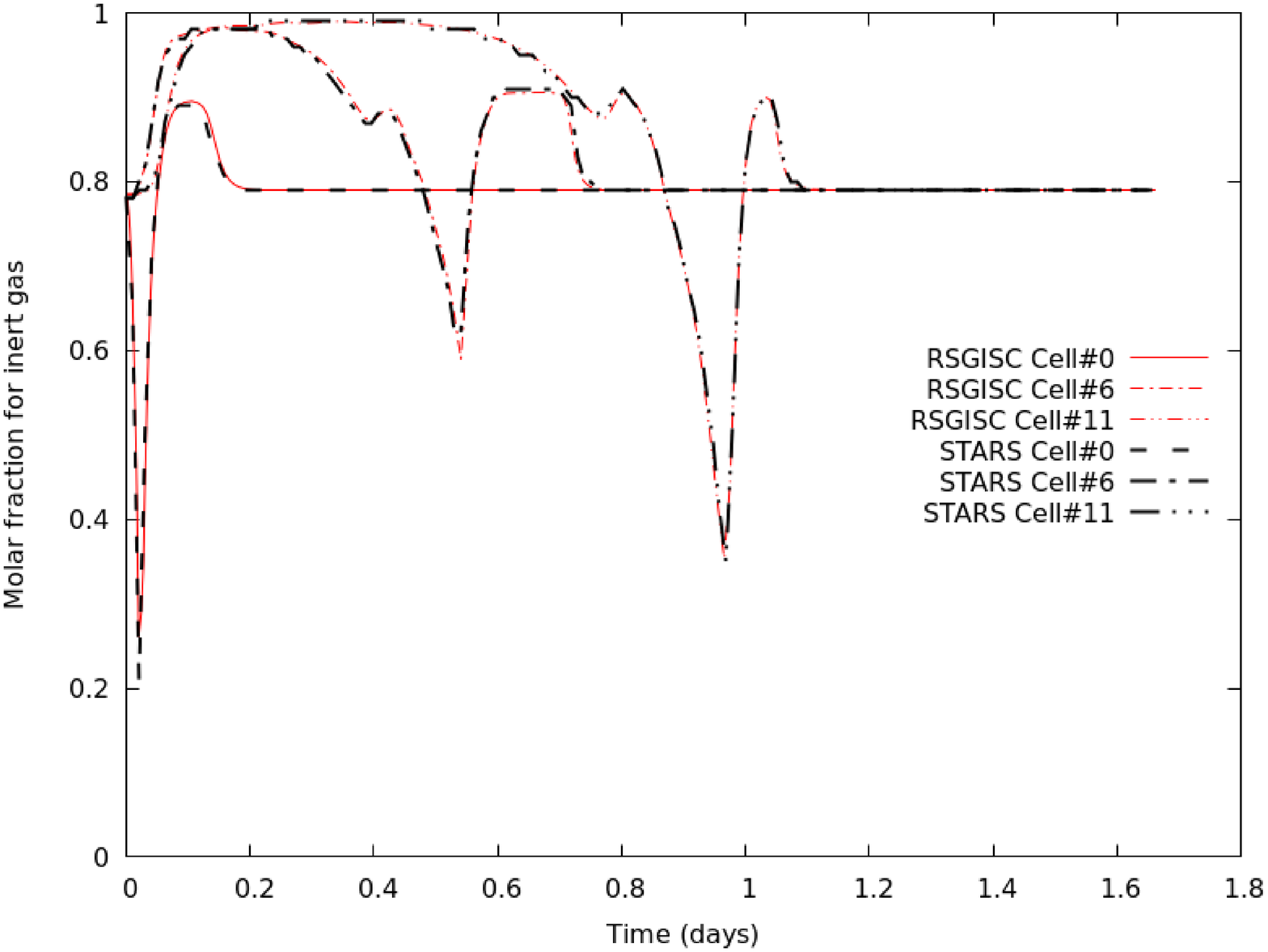}
 \caption{Combustion Tube Experiment: Gas molar fraction of inert gas profile with results from CMG STARS.}
 \label{fig:ex1_cmg_y4}
\end{figure}

Generally, some properties of three cells (\#0, 6, 11) are selected to compare.  From Figures
\ref{fig:ex1_cmg_t} to \ref{fig:ex1_cmg_y4}, the profiles of cell properties are presented, including
temperature, water saturation, oil saturation, the oil molar fraction of light oil, the gas molar fraction of
steam, oxygen and inert gas.  Also, well bottom hole pressures are compared with those from STARS.  From the
following figures, it indicates that the results from our simulator RSGISC match very well with those from CMG
STARS.
The properties mentioned here are all the primary unknowns in our simulator, and if these unknowns are
validated, all other properties can be obtained by these unknowns. Hence the validated unknowns ensure the
validity of other properties.

\subsection{Field-scale Simulation}

After a combustion tube experiment, a field-scale model is the task of the next step. We adopt the data from another set of data from different papers (\cite{crookston1979numerical}; \cite{grabowski1979fully}; \cite{vaughn1986numerical}). Historically used as a benchmark, this model is used to explain the validation of our simulator as well.

\subsubsection{Model Description}

\begin{figure}[!htb]
 \centering\includegraphics[width=0.75\textwidth]{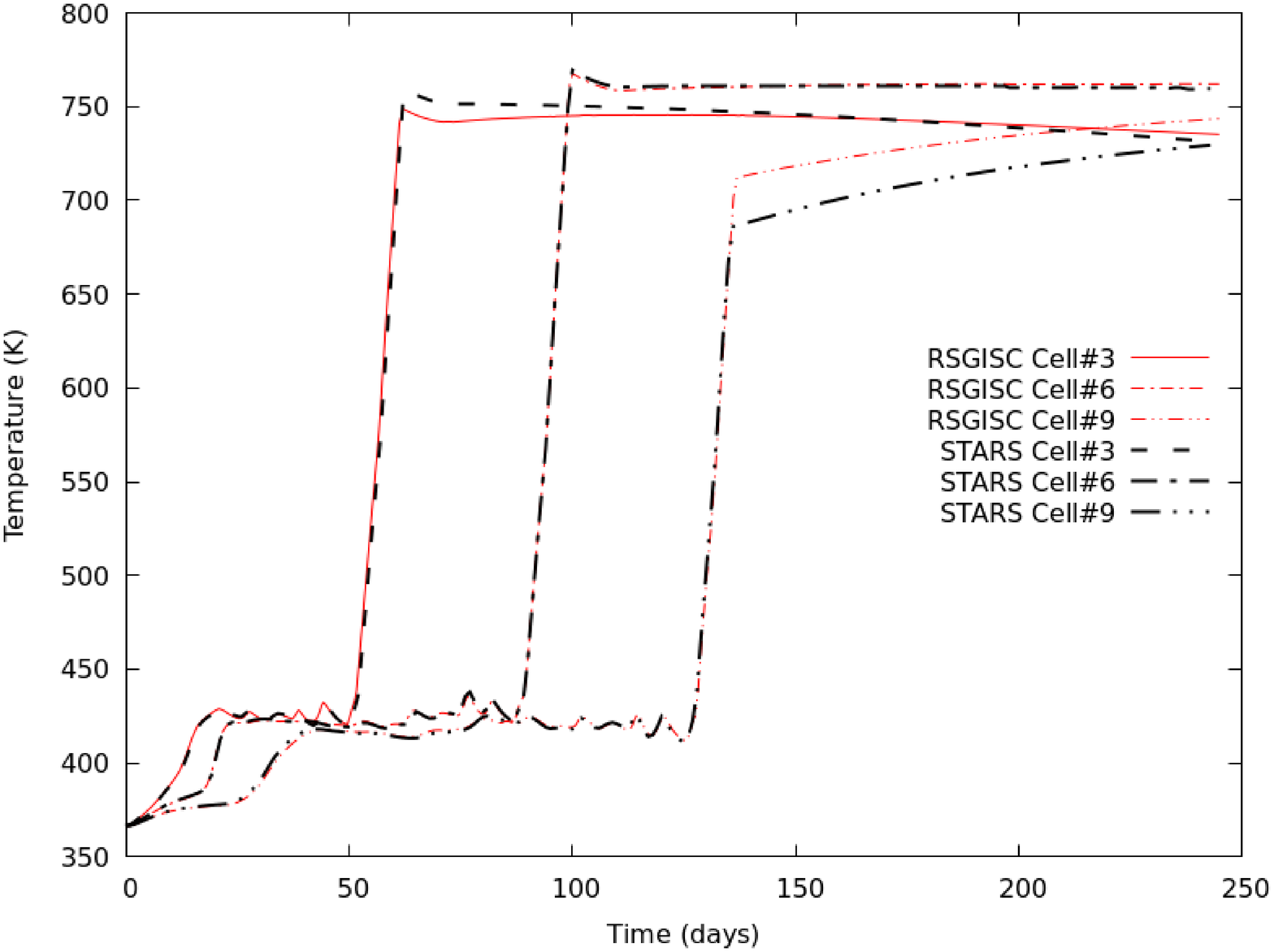}
 \caption{Field-scale: Temperature profile for field-scale model with results from CMG STARS.}
 \label{fig:ex2_cmg_t}
\end{figure}

\begin{figure}[!htb]
 \centering\includegraphics[width=0.75\textwidth]{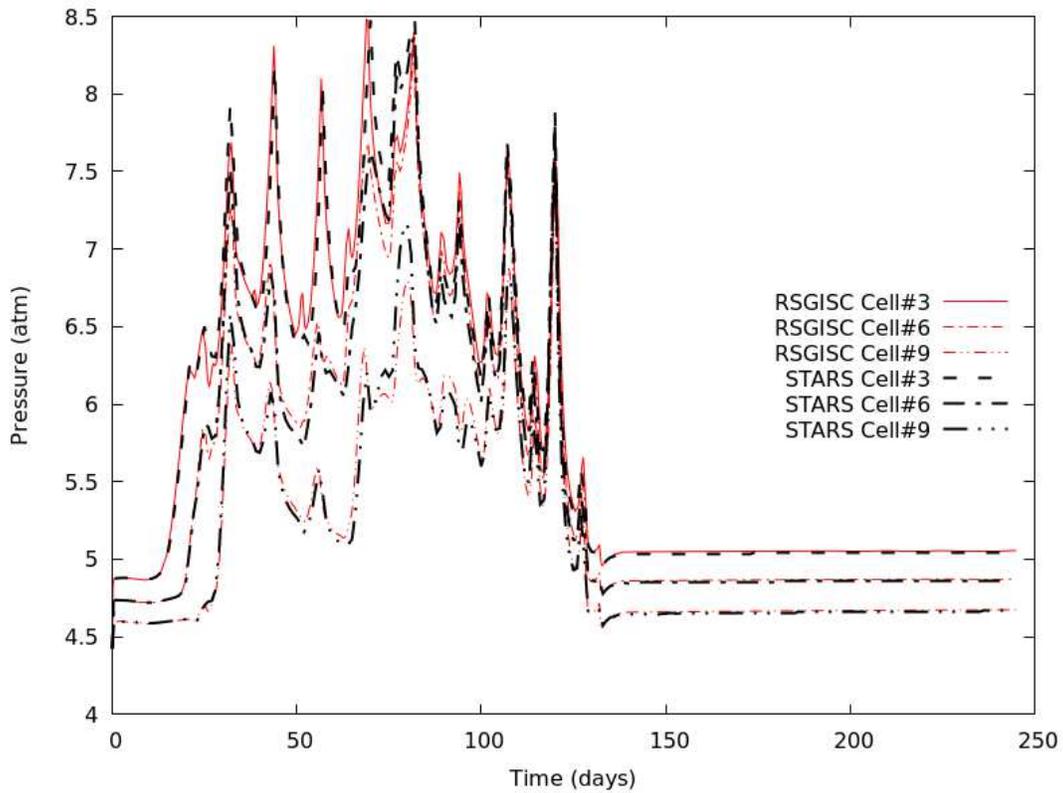}
 \caption{Field-scale: Pressure profile for field-scale model with results from CMG STARS.}
 \label{fig:ex2_cmg_p}
\end{figure}

\begin{figure}[!htb]
 \centering\includegraphics[width=0.75\textwidth]{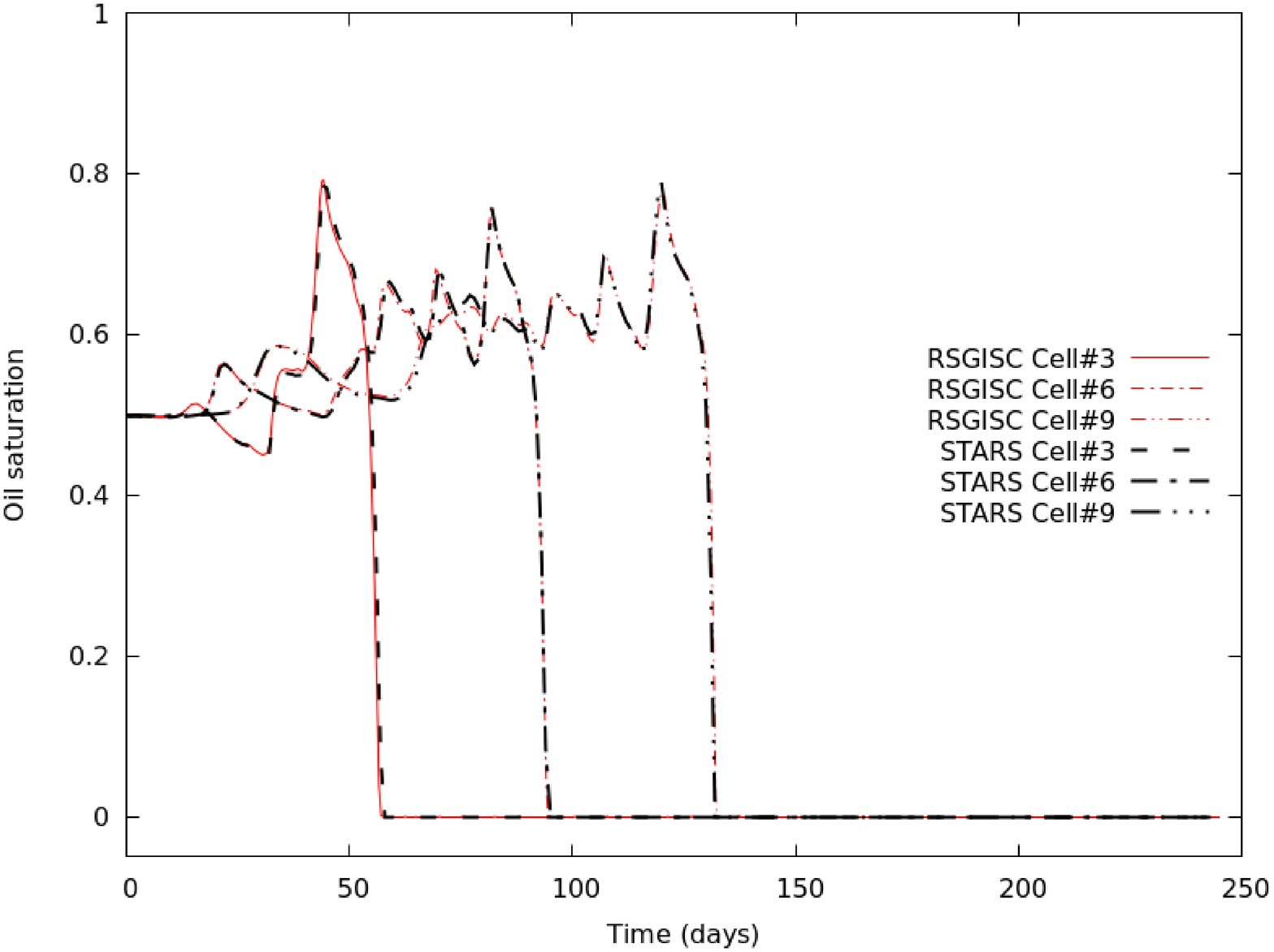}
 \caption{Field-scale: Oil saturation profile for field-scale model with results from CMG STARS.}
 \label{fig:ex2_cmg_so}
\end{figure}

\begin{figure}[!htb]
 \centering\includegraphics[width=0.75\textwidth]{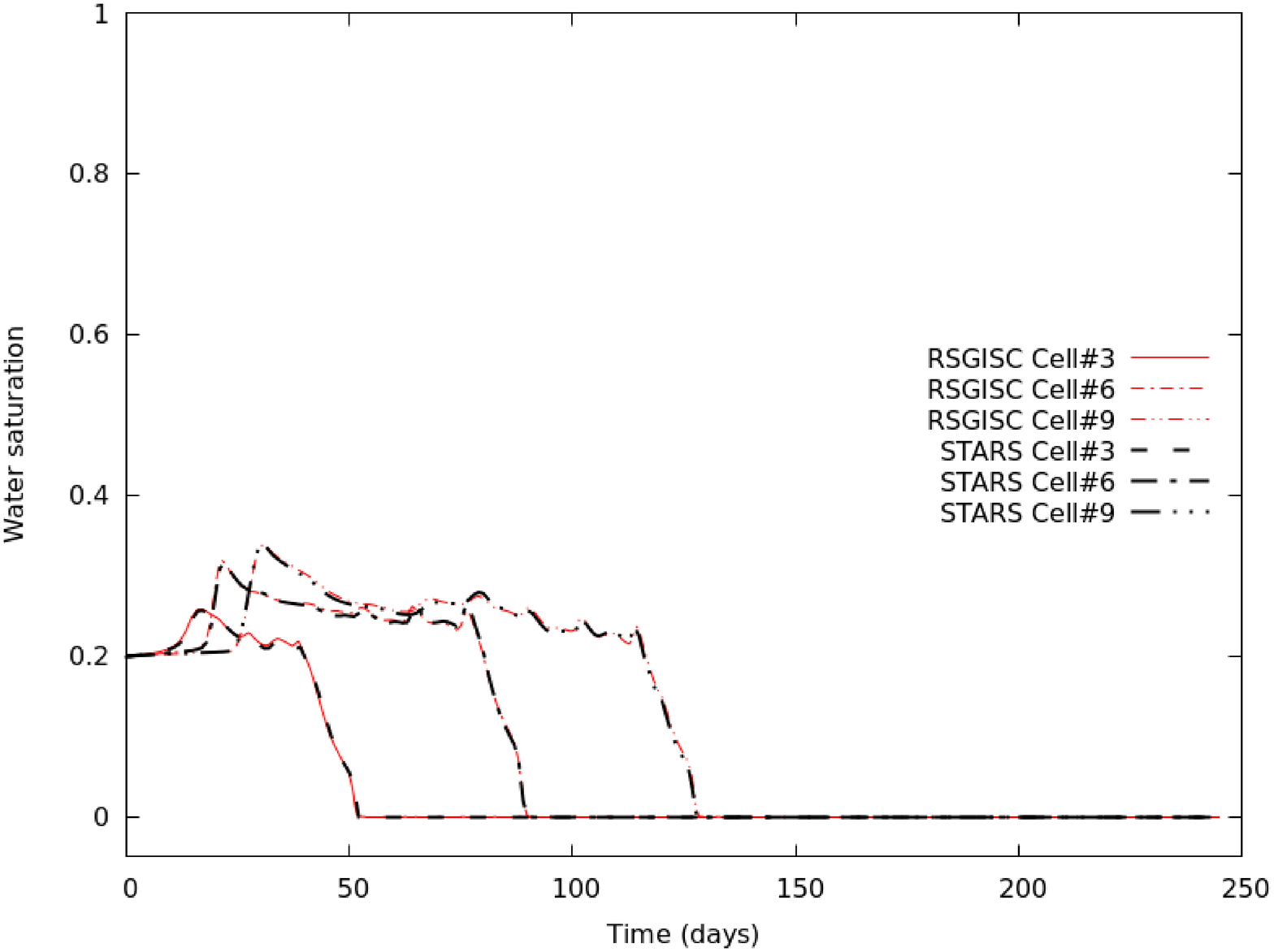}
 \caption{Field-scale: Water saturation profile for field-scale model with results from CMG STARS.}
 \label{fig:ex2_cmg_sw}
\end{figure}

\begin{figure}[!htb]
 \centering\includegraphics[width=0.75\textwidth]{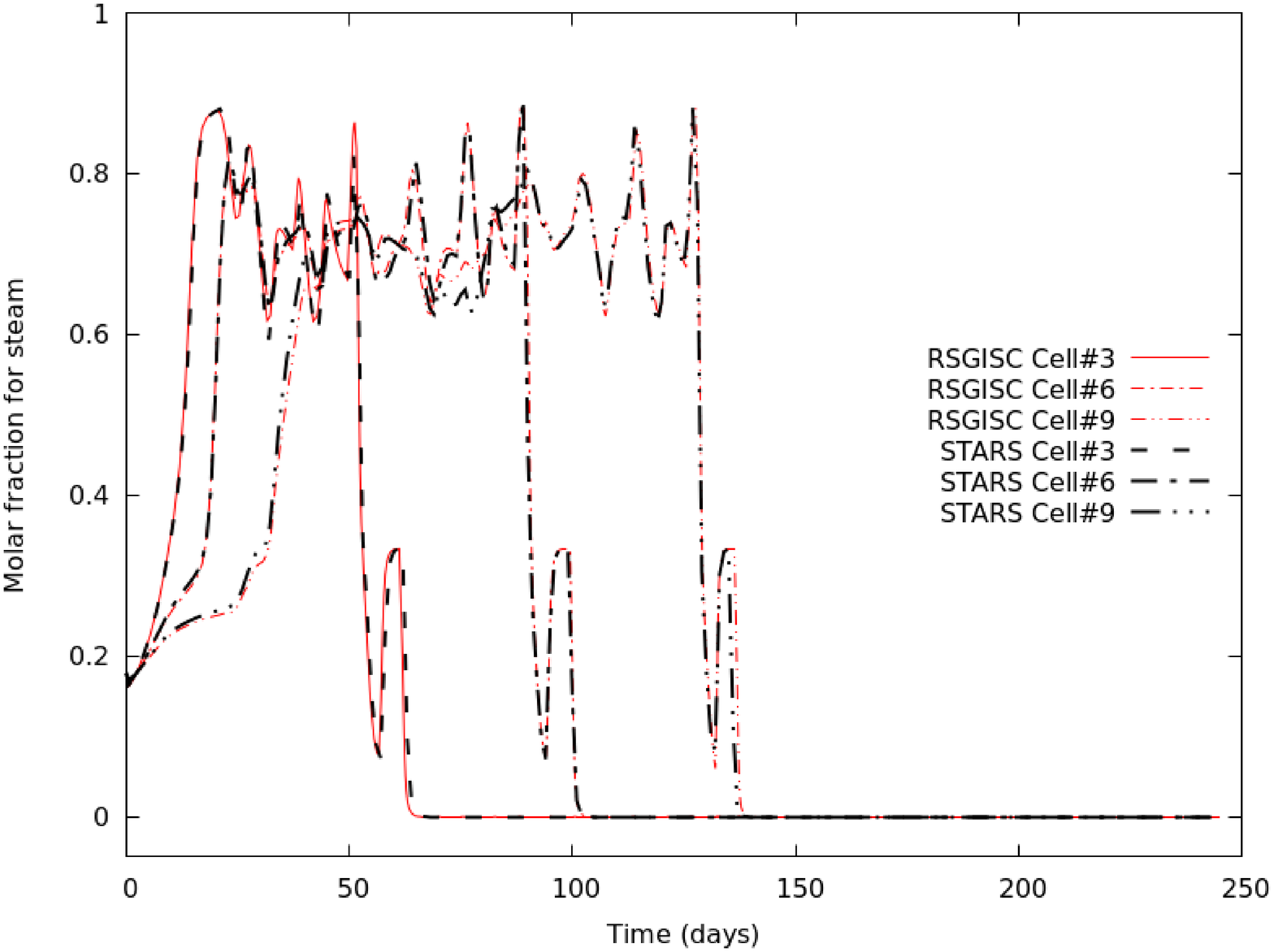}
 \caption{Field-scale: Gas molar fraction profile of steam for field-scale model with results from CMG STARS.}
 \label{fig:ex2_cmg_y0}
\end{figure}

The reservoir itself is 164ft long, 115ft wide and 21ft high, in which the behavior of in-situ combustion in
about one year is considered. In this model, pure oxygen is injected as the oxidizer. The oil has a density of
76.5 API at standard condition. The viscosity is 138.837 cp at initial condition.  The components and
corresponding reactions are different from the combustion tube model. All properties, reactions, and
operations are listed as follows. All denotations remain the same meanings as the previous model.
Also, the validations of some physical properties with CMG STARS are displayed in 
Figures \ref{fig:ex2_cmg_t} to \ref{fig:ex2_cmg_y0}, in which the temperature, pressure, saturations, molar
fractions are compared in certain cells. From these the figures, we can see that the match is excellent, which
confirms our simulator has correct results.

\begin{table}[!htb]
\centering
\begin{tabular}{l c}
\hline
\textbf{Initial condition} &  \\
$k_{x,y,z} (md)$  & 4000, 4000, 4000 \\
$\phi$ & 0.38 \\
$p (psi)$ & 65 \\
$T ({^\circ F})$ & 200 \\
$S_{w, o, g}$ & 0.2, 0.5, 0.3 \\
$x_{LO, HO}$ & 0.0 1.0\\
$y_{W, LO, HO, O2, IR}$ & 0.0, 0.0, 0.0, 1.0, 0.0\\
$C_c$ & 0 \\
\hline
\end{tabular}
\caption{Input data for field-scale model.}
\label{Tab:in1}
\end{table}

\begin{table}[!htb]
\centering
\begin{tabular}{l c c c c c c}
\hline
\textbf{Properties} & \textbf{H2O} & \textbf{LO} & \textbf{HO} & \textbf{O2} & \textbf{IR} & \textbf{Coke} \\
$M\ (lb/lbmole)$  & 18 & 44 & 170 & 32 & 44 & 13 \\
$p_{crit}\ (psi) $  & 3206.2 & 615.9 & 264.6 & 730 & 1073 &  \\
$T_{crit}\ ({}^\circ F)$  & 705.73 & 205.93 & 725.23 & -181.77 & 88.03 &  \\
\hline

$\rho_{ref} (lbmole/ft^3)$  & 3.47 & 0.4545 & 0.25 &   &   & 80.002 \\
$cp\ (1/psi)$  & 1.00e-5 & 7.69e-4 & 3.80e-4 &   &   &   \\
$ct1\ (1/{^\circ F})$  & 3.80e-4 & 2.20e-4 & 1.00e-5 &   &   & 3.9 \\
\hline

$cpg1\ (Btu/({^\circ F} \cdot lbmol))$  & 7.613 & -52.0192 & 57.8 & 7.68 & 11 &  \\
$cpg2\ (Btu/({^\circ F}^2 \cdot lbmol))$  & 8.616e-4 & 1.519e-1 & 6.018e-2 & 0 & 0 &  \\
$hvr\ (Btu/({^\circ F}^{ev} \cdot lbmol))$  & 1657 & 1917 & 12198 &  &  &  \\
$ev$  & 0.38 & 0.38 & 0.38 &  &  &  \\
\hline

$avg\ (cp/{^\circ F})$ & 8.822e-6 & 2.166e-6 & 3.926e-6 & 2.196e-4 & 2.127e-4\\
$bvg$ & 1.116 & 0.943 & 1.102 & 0.721 & 0.702\\
$avisc\ (cp)$ & 4.7352e-3 & 4.02e-4 & 4.02e-4 & & & \\
$bvisc\ ({^\circ F})$ & 2728.2 & 6121.6 & 6121.6 & & & \\
\hline

$kv1\ (psi)$ & 1.7202e6 & 1.4546e5 & 2.7454e5 & & & \\
$kv2\ (1/psi)$ & 0 & 0 & 0 & & & \\
$kv3$ & 0 & 0 & 0 & & & \\
$kv4\ ({^\circ F})$ & -6869.59 & -4458.73 & -8424.83 & & & \\
$kv5\ ({^\circ F})$ & -376.64 & -387.78 & -205.69 & & & \\
\hline
\end{tabular}
\caption{Input data for field-scale model (cont'd).}
\label{Tab:in2}
\end{table}

\begin{table}[!htb]
\centering
\begin{tabular}{l c}
\hline
\textbf{Reaction data} & \\
$LO + 5 O_2 \longrightarrow 3 IR + 4 H_2 O$ & $r_1$ \\
$HO + 18 O_2 \longrightarrow 11.64 IR + 13 H_2 O$ & $r_2$ \\
$HO \longrightarrow 2 LO + 4.67 Coke + 0.484 IR$ & $r_3$ \\
$Coke + 1.25 O_2 \longrightarrow 1 IR + 0.5 H_2 O$ & $r_4$\\
$A_R\ (1/(psi \cdot day))$ & 1e6, 1e6, 0.3e6, 1e6\\
$E_{aR}\ (Btu/lbmol)$ & 3.33e4, 3.33e4, 2.88e4, 2.34e4\\
$H_R\ (Btu/lbmol)$ & 9.48e5, 3.49e6, 2.0e4, 2.25e5\\
\end{tabular}
\\
\begin{tabular}{c c c}
\hline
\textbf{Well conditions} & & \\
Injector & gas rate (O2:N2 = 1:0) $(ft^3/hr)$ & 115000  \\
 & rc $(ft)$ & 0.5 \\
 & pinjw $(psi)$ & 10000 \\
 & tinjw $(^\circ F)$ & 200 \\
Producer & bhp $(psi)$ & 65 \\
\hline
\end{tabular}
\caption{Input data for field-scale model (cont'd).}
\label{Tab:in3}
\end{table}

\subsection{Scalability}

In this section, we adopt finer grids for in-situ combustion, and parallelization is implemented to accelerate
these models. For hardware, two clusters from Compute Canada are employed to our simulator, Niagara and
Graham.

For the 1D grid, grid dimenson is $1000000 \times 1 \times 1$. For the 2D grid, the grid dimension is $1000
\times 1000 \times 1$. And for the 3D grid, the grid dimension is $100 \times 100 \times 100$. For the three
models, 16 processors are used to handle the whole model, and 16 sub-grids are generated in the process of
grid partitioning.

The following part will consider the performance of some examples, and the scalability is the main factor of
concern. Also, some numerical treatments and grid characteristics are all taken into consideration.
The first case is employed as an in-situ combustion model. The grid is a sufficiently fine grid. The grid
dimension of the model is $1,000,000 \times 1 \times 1$ cells. That's actually a 1D model with one million
cells. The chemical model is the basic chemical model from (\cite{crookston1979numerical}). The BiCGSTAB
solver is employed with the tolerance of 1e-5. In addition, the stopping criterion for Newton iteration is
1e-2. The preconditioner pc$\_$index 3 is used.
The second and third models are applied with the different grids. For the second case, it is a 2D model, and
the grid dimension of the model is $1,000 \times 1,000 \times 1$ cells. For the third case, the 3D model, the
grid dimension is $100 \times 100 \times 100$.

\begin{example}
    \label{ex1}
    In this example, we would like to investigate the influence of nodes number. Bandwidth of memory access played
    an important role in the speedup profile.
    For the model, we directly apply the 3D model in last example, and run the simulation on 1 node, 2 nodes and 4
    nodes separately.  The numerical summary is listed below in Table \ref{Tab:para_example4},
    \ref{Tab:para_example5}, and \ref{Tab:para_example6}.
\end{example}

\begin{table}[!htb]
\centering
\begin{tabular}{|l|cccccc|}
\hline
\textbf{\#Processors} & 1 $\times$ 1 & 1 $\times$ 2 & 1 $\times$ 4 & 1 $\times$ 8 & 1 $\times$ 16 & 1 $\times$ 32\\\hline
\textbf{\#Steps} & 10 & 10 & 10 & 10 & 10 & 10 \\\hline
\textbf{\#Newton} & 22 & 22 & 22 & 22 & 22 & 22\\\hline
\textbf{\#Solver} & 22 & 23 & 26 & 26 & 28 & 33 \\\hline
\textbf{\#Avg. solver} & 1.00  & 1.05  & 1.18  & 1.18  & 1.27  & 1.50  \\\hline
\textbf{Time (s)} & 5026.9  & 2687.9  & 1364.3  & 733.7  & 416.6  & 243.9  \\\hline
\textbf{Avg. time (s)} & 228.50  & 116.87  & 52.47  & 28.22  & 14.88  & 7.39  \\\hline
\textbf{speedup (total)} & 1.000  & 1.870  & 3.684  & 6.851  & 12.066  & 20.614  \\\hline
\textbf{speedup (solver)} & 1.000  & 1.955  & 4.354  & 8.097  & 15.357  & 30.921  \\\hline
\end{tabular}
    \caption{Numerical summaries of 1-node case in Example \ref{ex1}.}
\label{Tab:para_example4}
\end{table}

\begin{table}[!htb]
\centering
\begin{tabular}{|l|cccccc|}
\hline
\textbf{\#Processors} & 2 $\times$ 1 & 2 $\times$ 2 & 2 $\times$ 4 & 2 $\times$ 8 & 2 $\times$ 16 & 2 $\times$ 32\\\hline
\textbf{\#Steps} & 10 & 10 & 10 & 10 & 10 & 10 \\\hline
\textbf{\#Newton} & 22 & 22 & 22 & 22 & 22 & 22\\\hline
\textbf{\#Solver} & 22 & 23 & 26 & 28 & 33 & 33 \\\hline
\textbf{\#Avg. solver} & 1.00 & 1.05 & 1.18 & 1.27 & 1.50 & 1.50 \\\hline
\textbf{Time (s)} & 2642.0 & 1356.8 & 723.8 & 395.9 & 232.1 & 140.9  \\\hline
\textbf{Avg. time (s)} & 120.09 & 58.99 & 27.84 & 14.14 & 7.03 & 4.27  \\\hline
\textbf{speedup (total)} & 1.000 & 1.947 & 3.650 & 6.673 & 11.385 & 18.746  \\\hline
\textbf{speedup (solver)} & 1.000 & 2.036 & 4.314 & 8.493 & 17.077 & 28.119  \\\hline
\end{tabular}
    \caption{Numerical summaries of 2-node case in Example \ref{ex1}.}
\label{Tab:para_example5}
\end{table}

\begin{table}[!htb]
\centering
\begin{tabular}{|l|cccccc|}
\hline
\textbf{\#Processors} & 4 $\times$ 1 & 4 $\times$ 2 & 4 $\times$ 4 & 4 $\times$ 8 & 4 $\times$ 16 & 4 $\times$ 32\\\hline
\textbf{\#Steps} & 10 & 10 & 10 & 10 & 10 & 10 \\\hline
\textbf{\#Newton} & 22 & 22 & 22 & 22 & 22 & 22 \\\hline
\textbf{\#Solver} & 23 & 26 & 28 & 33 & 35 & 36 \\\hline
\textbf{\#Avg. solver} & 1.05  & 1.18  & 1.27  & 1.50  & 1.59  & 1.64  \\\hline
\textbf{Time (s)} & 1330.2  & 717.9  & 390.4  & 220.3  & 129.4  & 78.4  \\\hline
\textbf{Avg. time (s)} & 57.84  & 27.61  & 13.94  & 6.68  & 3.70  & 2.18  \\\hline
\textbf{speedup (total)} & 1.000  & 1.853  & 3.408  & 6.039  & 10.277  & 16.960  \\\hline
\textbf{speedup (solver)} & 1.000  & 2.095  & 4.149  & 8.664  & 15.639  & 26.547  \\\hline
\end{tabular}
    \caption{Numerical summaries of 4-node case in Example \ref{ex1}.}
\label{Tab:para_example6}
\end{table}

\begin{figure}[!htb]
 \centering\includegraphics[width=0.75\textwidth]{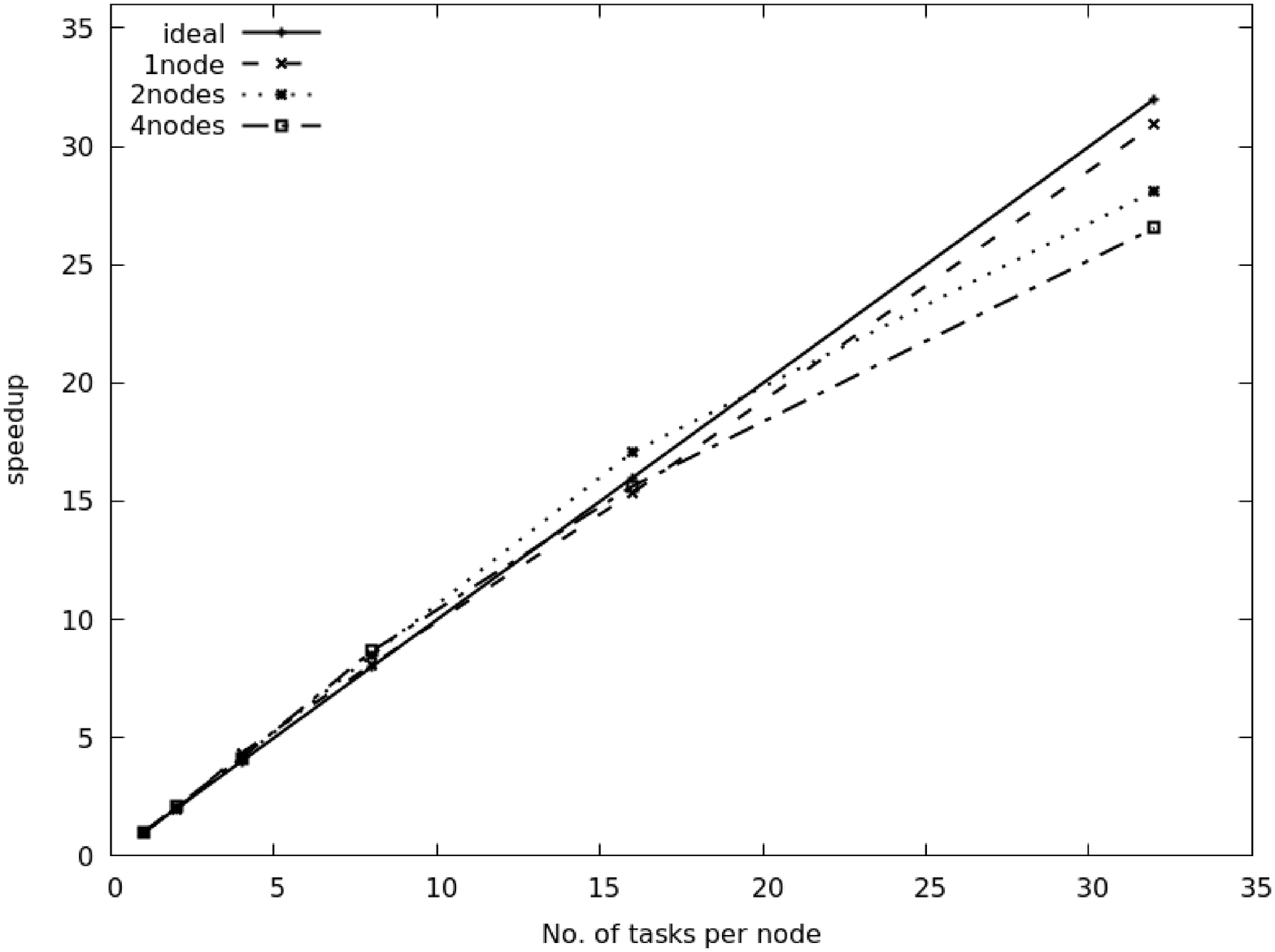}
    \caption{Comparison of speedup profiles of Example \ref{ex1}.}
 \label{fig:para_compare_acc2}
\end{figure}

From the numerical summaries, it can be noticed that number of solver increases rapidly with number of tasks
per node increasing, which has a significant effect on the speedup. Thus, a speedup of solver is chosen for
comparison.
In Figure \ref{fig:para_compare_acc2}, it indicates that number of nodes will affect the effects the result of
speedup. Generally, when the number of nodes increases, more data communications are required, which will cost
a large number of time. In the 1-nodes case, all data communications happen between cores on the same node.
Message passing between cores on one node is faster than that on different nodes.

\begin{figure}[!htb]
 \centering\includegraphics[width=0.75\textwidth]{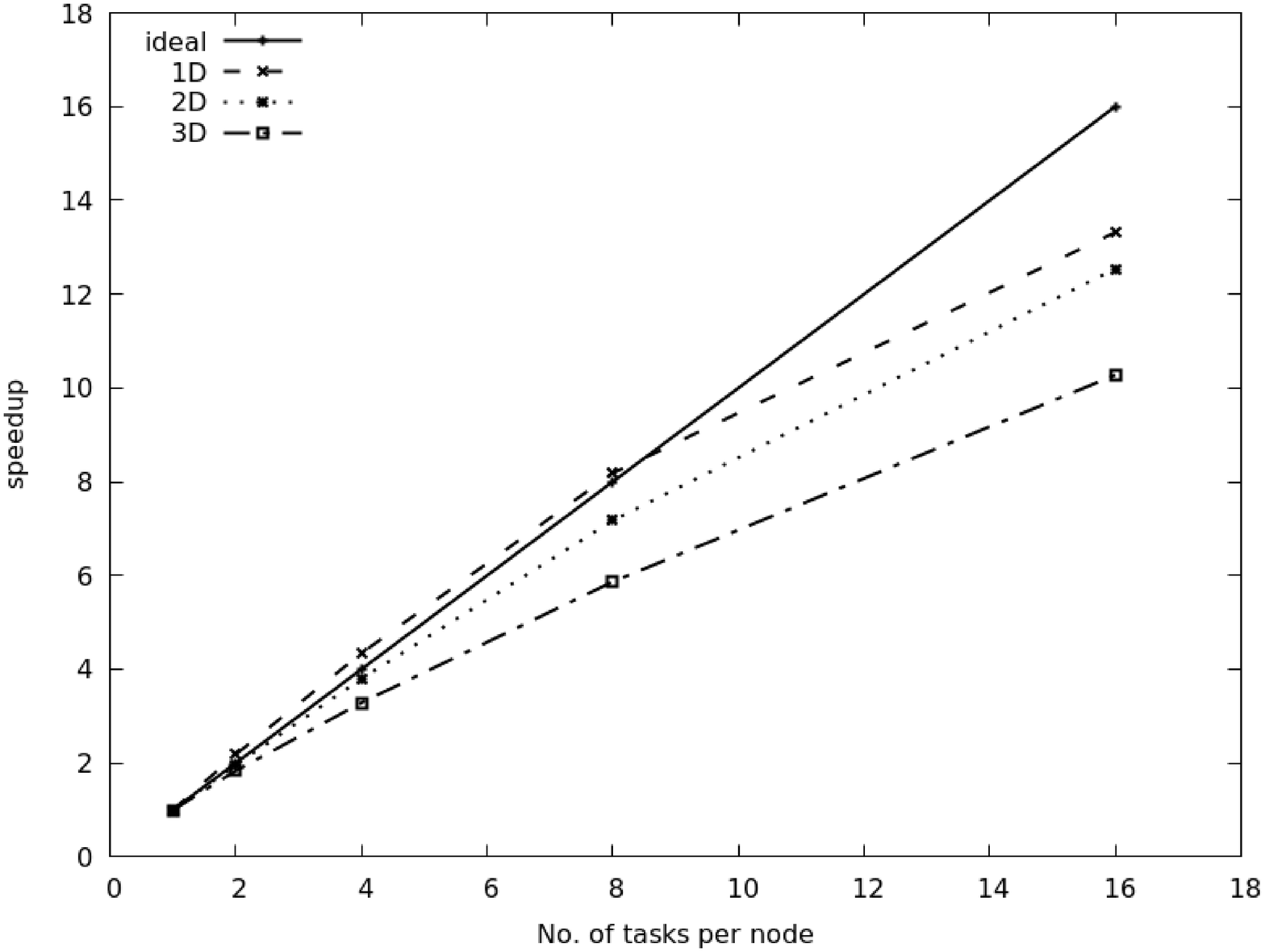}
    \caption{Comparison of speedup profiles of Example \ref{ex2}.}
 \label{fig:para_compare_acc3}
\end{figure}

\begin{example}
    \label{ex2}
    In this section, we adopt the models in previous example, and changes the tolerance of linear solver. Also, the time
    steps are adjusted to similar. Preconditioner is changed to CPR-PF from RAS. With this modification, the
    number of solvers will remain stable when number of tasks increases. Also, for efficiency of parallelization,
    4 nodes are adopted in this example.
    The numerical summary is listed below in Table \ref{Tab:para_example7}, \ref{Tab:para_example8} and
    \ref{Tab:para_example9}.
    It indicates from Figure \ref{fig:para_compare_acc3} that without the effects of unstable solver, the speedup
    is better.
\end{example}

\begin{table}[!htb]
\centering
\begin{tabular}{|l|ccccc|}
\hline
\textbf{\#Processors} & 4 $\times$ 1 & 4 $\times$ 2 & 4 $\times$ 4 & 4 $\times$ 8 & 4 $\times$ 16\\\hline
\textbf{\#Steps} & 41 & 41 & 41 & 41 & 41 \\\hline
\textbf{\#Newton} & 77 & 77 & 77 & 77 & 77 \\\hline
\textbf{\#Solver} & 77 & 77 & 77 & 77 & 77 \\\hline
\textbf{\#Avg. solver} & 1.00 & 1.00 & 1.00 & 1.00 & 1.00 \\\hline
\textbf{Time (s)} & 2248.2 & 1029.2 & 518.0 & 274.7 & 168.8 \\\hline
\textbf{Avg. time (s)} & 29.197 & 13.366 & 6.728 & 3.568 & 2.192 \\\hline
\textbf{speedup (total)} & 1.000 & 2.184 & 4.340 & 8.183 & 13.319 \\\hline
\textbf{speedup (solver)} & 1.000 & 2.184 & 4.340 & 8.183 & 13.319 \\\hline
\end{tabular}
    \caption{Numerical summaries of 1D case in Example \ref{ex2}.}
\label{Tab:para_example7}
\end{table}

\begin{table}[!htb]
\centering
\begin{tabular}{|l|ccccc|}
\hline
\textbf{\#Processors} & 4 $\times$ 1 & 4 $\times$ 2 & 4 $\times$ 4 & 4 $\times$ 8 & 4 $\times$ 16\\\hline
\textbf{\#Steps} & 48 & 48 & 48 & 48 & 48 \\\hline
\textbf{\#Newton} & 78 & 78 & 78 & 78 & 78 \\\hline
\textbf{\#Solver} & 78 & 78 & 78 & 78 & 78 \\\hline
\textbf{\#Avg. solver} & 1.00 & 1.00 & 1.00 & 1.00 & 1.00 \\\hline
\textbf{Time (s)} & 2900.4 & 1481.9 & 764.0 & 404.5 & 231.2 \\\hline
\textbf{Avg. time (s)} & 37.184 & 18.999 & 9.794 & 5.185 & 2.965 \\\hline
\textbf{speedup (total)} & 1.000 & 1.957 & 3.797 & 7.171 & 12.543 \\\hline
\textbf{speedup (solver)} & 1.000 & 1.957 & 3.797 & 7.171 & 12.543 \\\hline
\end{tabular}
    \caption{Numerical summaries of 2D case in Example \ref{ex2}.}
\label{Tab:para_example8}
\end{table}

\begin{table}[!htb]
\centering
\begin{tabular}{|l|ccccc|}
\hline
\textbf{\#Processors} & 4 $\times$ 1 & 4 $\times$ 2 & 4 $\times$ 4 & 4 $\times$ 8 & 4 $\times$ 16\\\hline
\textbf{\#Steps} & 60 & 60 & 60 & 60 & 60 \\\hline
\textbf{\#Newton} & 72 & 72 & 72 & 72 & 72 \\\hline
\textbf{\#Solver} & 72 & 72 & 72 & 75 & 83 \\\hline
\textbf{\#Avg. solver} & 1.00 & 1.00 & 1.00 & 1.04 & 1.15 \\\hline
\textbf{Time (s)} & 4716.0 & 2568.1 & 1434.6 & 803.9 & 458.9 \\\hline
\textbf{Avg. time (s)} & 65.499 & 35.668 & 19.926 & 10.718 & 5.529 \\\hline
\textbf{speedup (total)} & 1.000 & 1.836 & 3.287 & 5.867 & 10.277 \\\hline
\textbf{speedup (solver)} & 1.000 & 1.836 & 3.287 & 6.111 & 11.847 \\\hline
\end{tabular}
    \caption{Numerical summaries of 3D case in Example \ref{ex2}.}
\label{Tab:para_example9}
\end{table}

\begin{figure}[!htb]
 \centering\includegraphics[width=0.5\textwidth, angle=270]{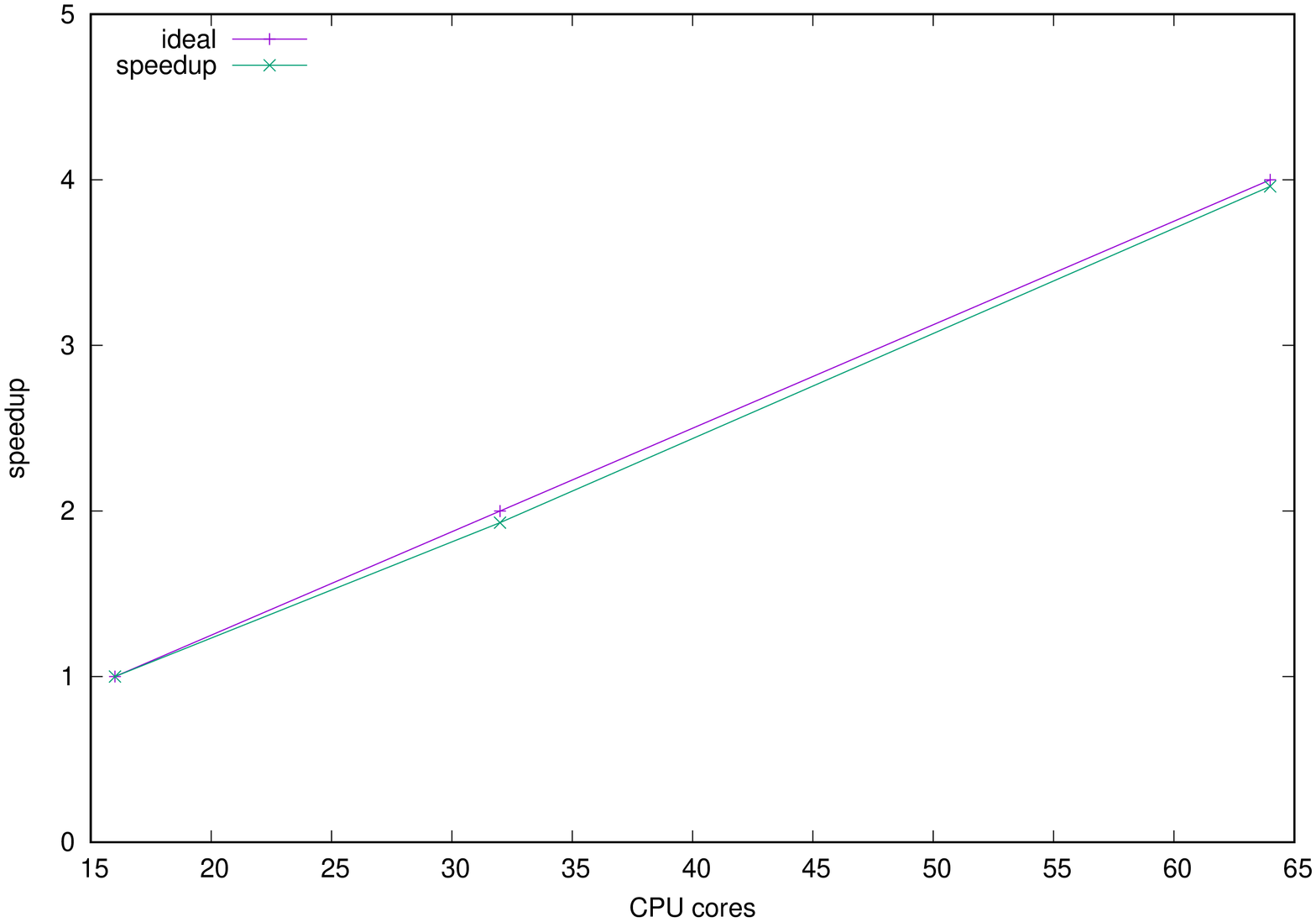}
    \caption{Comparison of speedup profiles of Example \ref{ex3}.}
 \label{fig:para_compare_acc1}
\end{figure}

\begin{example}
    \label{ex3}
    This example considers a model with grid dimension $360 \times 200 \times 160$, which has around 11.5
    million grid blocks. Up to 64 CPU cores are employed. 
    Its scalability is shown by Figure \ref{fig:para_compare_acc1}, from which we can see that the scalability
    is excellent.
\end{example}

\section{Conclusion}

In this work, a new parallel In-Situ Combustion Simulator is developed. On the mathematical model, some
modifications of methods are implemented. For numerical methods, a novel decoupling method is proposed. In
parallelization, some advanced algorithms are first adopted to an ISC simulator.

For the mathematical base of the simulator, both mass conservation equations and energy conservation equations
are included. With well equations and the oil phase and gas phase constraint equations, a PDE (partial
differential equation) system is established. The multi-component multi-phase flow is described by
conservation equations and Darcy's flow equations. Effects of temperature, pressure, and saturation on fluids,
rock and rock-fluid properties are taken into account. With all these equations, the non-isothermal process
can be well described. For chemical reactions, the Arrhenius models are considered to couple with the
equations. With the mathematical model, the discretization and linearization can be achieved with the finite
difference method and Newton method. For a general component model, different from adopting the VS method, a
modified PER method is adopted. The results indicate its simplicity and effectiveness. In the validation part,
the total error of the simulator is not apparent, which illustrates that the error introduced by this method
is tiny. There is no noticeable difference between this method and the VS method in the aspect of accuracy.
In the numerical method, a novel decoupling method, Gauss-Jordan transformation is introduced. Due to the
existence of multiple components, the strong coupling phenomenon resulted in the complexity of solving an ISC
problem. Without a proper decoupling method, the parallel iterative linear solver cannot reach convergence in
expected steps. The new decoupling method not only solves the coupling phenomenon, but its efficiency is also
better than some traditional decoupling methods. Furthermore, with a row exchange, a dynamic equation
alignment is achieved, which is suitable for the process with rapid and frequent phase appearance and
disappearance.

With the simulator, results can be obtained from a lab-scale model to a field-scale model. Results from the
multi-dimensional model can be validated with CMG STARS, and properties and profiles match the physical
process as well. Both suggest positive validations of our simulator. Afterward, the performance of
parallelization is evaluated, which indicates the scalability of the simulator.

\section*{Acknowledgements}
This work is partially supported by Department of Chemical Petroleum Engineering, 
University of Calgary, NSERC, AIEES, Foundation CMG, AITF iCore, IBM Thomas J. Watson 
Research Center, Frank and Sarah Meyer FCMG Collaboration Center, 
WestGrid (www.westgrid.ca), 
SciNet (www.scinetpc.ca) and Compute Canada Calcul Canada (www.computecanada.ca).


\begin{thebibliography}{10}

\bibitem{abou1985handling}
{Abou-Kassem, J. H.} and {Aziz, K.}
\newblock Handling of phase change in thermal simulators.
\newblock {\em Journal of petroleum technology}, 37(09):1661--1663, 1985.

\bibitem{adler1975linear}
{Adler, G.}
\newblock A linear model and a related very stable numerical method for thermal
  secondary oil recovery.
\newblock {\em Journal of Canadian Petroleum Technology}, 14(03), 1975.

\bibitem{akkutlu2003dynamics}
{Akkutlu, I. Y.} and {Yortsos, Y. C.}
\newblock The dynamics of in-situ combustion fronts in porous media.
\newblock {\em Combustion and Flame}, 134(3):229--247, 2003.

\bibitem{ali1977multiphase}
{Ali, S. M.}
\newblock Multiphase, multidimensional simulation of in situ combustion.
\newblock In {\em SPE Annual Fall Technical Conference and Exhibition}. Society
  of Petroleum Engineers, 1977.

\bibitem{anis1983sensitivity}
{Anis, M.}, {Hwang, M. K.}, and {Odeh, A. S.}
\newblock A sensitivity study of the effect of parameters on results from an
  in-situ combustion simulator.
\newblock {\em Society of Petroleum Engineers Journal}, 23(2):259--264, 1983.

\bibitem{bailey1959heat}
{Bailey, H. R.} and {Larkin, B. K.}
\newblock Heat conduction in underground combustion.
\newblock {\em Transactions of The American Institute of Mining and
  Metallurgical Engineers}, 216:123--129, 1959.

\bibitem{bailey1960conduction}
{Bailey, H. R.} and {Larkin, B. K.}
\newblock Conduction-convection in underground combustion.
\newblock {\em Petroleum Transactions, AIME}, 219:320--331, 1960.

\bibitem{bank1989alternate}
{Bank, R. E.}, {Chan, T. F.}, {Coughran, W. M.}, and {Smith, R. K.}
\newblock The alternate-block-factorization procedure for systems of partial
  differential equations.
\newblock {\em BIT Numerical Mathematics}, 29(4):938--954, 1989.

\bibitem{belgrave1993comprehensive}
{Belgrave, J. D. M.}, {Moore, R. G.}, {Ursenbach, M. G.}, and {Bennion, D. W.}
\newblock A comprehensive approach to in-situ combustion modeling.
\newblock {\em SPE Advanced Technology Series}, 1(01):98--107, 1993.

\bibitem{chen2014kinetic}
{Chen, X.}, {Chen, Z. J.}, {Moore, R. G.}, {Mehta, S. A.}, {Ursenbach, M. G.},
  and {Harding, T. G.}
\newblock Kinetic modeling of the in-situ combustion process for athabasca oil
  sands.
\newblock In {\em SPE Heavy Oil Conference-Canada}. Society of Petroleum
  Engineers, 2014.

\bibitem{chen2007reservoir}
{Chen, Z.}
\newblock {\em Reservoir Simulation: Mathematical Techniques in Oil Recovery}.
\newblock CBMS-NSF Regional Conference Series in Applied Mathematics. SIAM,
  2007.

\bibitem{chu1963two}
{Chu, C.}
\newblock Two-dimensional analysis of a radial heat wave.
\newblock {\em Journal of Petroleum Technology}, 15(10):1137--1144, 1963.

\bibitem{clapeyron1834memoir}
{Clapeyron, É.}
\newblock Mémoir sur la puissance motrice de la chaleur.
\newblock {\em Journal de l'École Royale Polytechnique}, pages 153--190, 1834.

\bibitem{cmg2015starguide}
{CMG}.
\newblock {\em STARS User's Guide}.
\newblock Computer Modelling Group Ltd., 2015.

\bibitem{cmgstars}
{CMG STARS}.
\newblock Thermal and advanced processes simulator.
\newblock https://www.cmgl.ca/stars, 2018.

\bibitem{coats1976simulation}
{Coats, K. H.}
\newblock Simulation of steamflooding with distillation and solution gas.
\newblock {\em Society of Petroleum Engineers Journal}, 16(05):235--247, 1976.

\bibitem{coats1980situ}
{Coats, K. H.}
\newblock In-situ combustion model.
\newblock {\em Society of Petroleum Engineers Journal}, 20(06):533--554, 1980.

\bibitem{coats1983some}
{Coats, K. H.}
\newblock Some observations on field-scale simulation of the in-situ combustion
  process.
\newblock In {\em SPE Reservoir Simulation Symposium}. Society of Petroleum
  Engineers, 1983.

\bibitem{conti2016international}
{Conti, J.}, {Holtberg, P.}, {Diefenderfer, J.}, {LaRose, A.}, {Turnure, J.
  T.}, and {Westfall, L.}
\newblock International energy outlook 2016 with projections to 2040.
\newblock Technical report, USDOE Energy Information Administration (EIA),
  Washington, DC (United States). Office of Energy Analysis, 2016.

\bibitem{corey1956three}
{Corey, A. T.}, {Rathjens, C. H.}, {Henderson, J. H.}, and {Wyllie, M. R. J.}
\newblock Three-phase relative permeability.
\newblock {\em Journal of Petroleum Technology}, 8(11):63--65, 1956.

\bibitem{crookston1979numerical}
{Crookston, R. B.}, {Culham, W. E.}, and {Chen, W. H.}
\newblock A numerical simulation model for thermal recovery processes.
\newblock {\em Society of Petroleum Engineers Journal}, 19(01):37--58, 1979.

\bibitem{delshad1989comparison}
{Delshad, M.} and {Pope, G. A.}
\newblock Comparison of the three-phase oil relative permeability models.
\newblock {\em Transport in Porous Media}, 4(1):59--83, 1989.

\bibitem{ertekin2001basic}
{Ertekin, T.}, {Abou-Kassem, J. H.}, and {King, G. R.}
\newblock {\em Basic Applied Reservoir Simulations}.
\newblock Society of Petroleum Engineers, 2001.

\bibitem{gottfried1965mathematical}
{Gottfried, B. S.}
\newblock A mathematical model of thermal oil recovery in linear systems.
\newblock {\em Society of Petroleum Engineers Journal}, 5(03):196--210, 1965.

\bibitem{grabowski1979fully}
{Grabowski, J. W.}, {Vinsome, P. K.}, {Lin, R. C.}, {Behie, G. A.}, and {Rubin,
  B.}
\newblock A fully implicit general purpose finite-difference thermal model for
  in-situ combustion and steam.
\newblock In {\em SPE Annual Technical Conference and Exhibition}. Society of
  Petroleum Engineers, 1979.

\bibitem{gries2015system}
{Gries, S.}
\newblock {\em System-AMG Approaches for Industrial Fully and Adaptive Implicit
  Oil Reservoir Simulations}.
\newblock PhD thesis, Universit{\"a}t zu K{\"o}ln, 2015.

\bibitem{gutierrez2012abcs}
{Gutierrez, D.}, {Moore, R. G.}, {Ursenbach, M. G.}, and {Mehta, S. A.}
\newblock The abcs of in-situ combustion simulations: from laboratory
  experiments to field scale.
\newblock {\em Journal of Canadian Petroleum Technology}, 51(04):256--267,
  2012.

\bibitem{hwang1982situ}
{Hwang, M. K.}, {Jines, W. R.}, and {Odeh, A. S.}
\newblock An in-situ combustion process simulator with a moving-front
  representation.
\newblock {\em Society of Petroleum Engineers Journal}, 22(02):271--279, 1982.

\bibitem{ito1988field}
{Ito, Y.} and {Chow, A. K. Y.}
\newblock A field scale in-situ combustion simulator with channeling
  considerations.
\newblock {\em SPE reservoir engineering}, 3(02):419--430, 1988.

\bibitem{jiang2008techniques}
{Jiang, Y.}
\newblock {\em Techniques for Modeling Complex Reservoirs and Advanced Wells}.
\newblock PhD thesis, Stanford University, 2008.

\bibitem{kelly2010oil}
{Kelly, E. N.}, {Schindler, D. W.}, {Hodson, P. V.}, {Short, J. W.},
  {Radmanovich, R.}, and {Nielsen, C. C.}
\newblock Oil sands development contributes elements toxic at low
  concentrations to the athabasca river and its tributaries.
\newblock {\em Proceedings of the National Academy of Sciences},
  107(37):16178--16183, 2010.

\bibitem{klie1996two}
{Klie, H.}, {Rame, M.}, and {Wheeler, M.}
\newblock Two-stage preconditions for inexact newton methods in multi-phase
  reservoir simulation.
\newblock Technical report, Center for Research on Parallel Computation, Rice
  University, 1996.

\bibitem{kristensen2008development}
{Kristensen, M. R.}
\newblock {\em Development of Models and Algorithms for the Study of Reactive
  Porous Media Processes}.
\newblock PhD thesis, 2008.

\bibitem{lacroix2000iterative}
{Lacroix, S.}, {Vassilevski, Y. V.}, and {Wheeler, M. F.}
\newblock Iterative solvers of the implicit parallel accurate reservoir
  simulator (ipars), i: Single processor case.
\newblock Technical report, TICAM, The University of Texas at Austin, 2000.

\bibitem{lin1984numerical}
{Lin, C. Y.}, {Chen, W. H.}, {Lee, S. T.}, and {Culham, W. E.}
\newblock Numerical simulation of combustion tube experiments and the
  associated kinetics of in-situ combustion processes.
\newblock {\em Society of Petroleum Engineers Journal}, 24(06):657--666, 1984.

\bibitem{liu2016family}
{Liu, H.}, {Wang, K.}, and {Chen, Z.}
\newblock A family of constrained pressure residual preconditioners for
  parallel reservoir simulations.
\newblock {\em Numerical linear algebra with applications}, 23(1):120--146,
  2016.

\bibitem{liu2015parallel}
{Liu, H.}, {Wang, K.}, {Chen, Z.}, {Jordan, K. E.}, {Luo, J.}, and {Deng, H.}
\newblock A parallel framework for reservoir simulators on distributed-memory
  supercomputers.
\newblock In {\em SPE/IATMI Asia Pacific Oil \& Gas Conference and Exhibition}.
  Society of Petroleum Engineers, 2015.

\bibitem{liu2017dynamic}
{Liu, H.}, {Wang, K.}, {Yang, B.}, {Yang, M.}, {He, R.}, {Shen, L.}, {Zhong,
  H.}, and {Chen, Z.}
\newblock Dynamic load balancing using hilbert space-filling curves for
  parallel reservoir simulations.
\newblock In {\em SPE Reservoir Simulation Conference}. Society of Petroleum
  Engineers, 2017.

\bibitem{liu2016performance}
{Liu, H.}, {Zhang, P.}, {Wang, K.}, {Yang, B.}, and {Chen, Z.}
\newblock Performance and scalability analysis for parallel reservoir
  simulations on three supercomputer architectures.
\newblock In {\em Proceedings of the XSEDE16 Conference on Diversity, Big Data,
  and Science at Scale}, page~9. ACM, 2016.

\bibitem{mahinpey2007situ}
{Mahinpey, N.}, {Ambalae, A.}, and {Asghari, K.}
\newblock In-situ combustion in enhanced oil recovery (eor): A review.
\newblock {\em Chemical Engineering Communications}, 194(8):995--1021, 2007.

\bibitem{moore1995situ}
{Moore, R. G.}, {Laureshen, C. J.}, {Belgrave, J. D.}, {Ursenbach, M. G.}, and
  {Mehta, S. R. }.
\newblock In-situ combustion in canadian heavy oil reservoirs.
\newblock {\em Fuel}, 74(8):1169--1175, 1995.

\bibitem{naar1961three}
{Naar, J.} and {Wygal, R. J.}
\newblock Three-phase imbibition relative permeability.
\newblock {\em Society of Petroleum Engineers Journal}, 1(04):254--258, 1961.

\bibitem{nissen2015upscaling}
{Nissen, A.}, {Zhu, Z.}, {Kovscek, A.}, {Castanier, L.}, and {Gerritsen, M.}
\newblock Upscaling kinetics for field-scale in-situ combustion simulation.
\newblock {\em SPE Reservoir Evaluation \& Engineering}, 18(02):158--170, 2015.

\bibitem{odeh1969reservoir}
{Odeh, A. S.}
\newblock Reservoir simulation... what is it.
\newblock {\em Journal of Petroleum technology}, 21(11):1383--1388, 1969.

\bibitem{oklany1992situ}
{Oklany, J. S. F.}
\newblock {\em An In-situ Combustion Simulator for Enhanced Oil Recovery}.
\newblock PhD thesis, University of Salford, 1992.

\bibitem{peaceman1978interpretation}
{Peaceman, D. W.}
\newblock Interpretation of well-block pressures in numerical reservoir
  simulation.
\newblock {\em Society of Petroleum Engineers Journal}, 18(03):183--194, 1978.

\bibitem{peng1976new}
{Peng, D. Y.} and {Robinson, D. B.}
\newblock A new two-constant equation of state.
\newblock {\em Industrial \& Engineering Chemistry Fundamentals}, 15(1):59--64,
  1976.

\bibitem{ramey1959transient}
{Ramey Jr, H. J.}
\newblock Transient heat conduction during radial movement of a cylindrical
  heat source-applications to the thermal recovery process.
\newblock {\em Petroleum Transactions, AIME}, 216:115--122, 1959.

\bibitem{redlich1949thermodynamics}
{Redlich, O.} and {Kwong, J. N.}
\newblock On the thermodynamics of solutions. v. an equation of state.
  fugacities of gaseous solutions.
\newblock {\em Chemical reviews}, 44(1):233--244, 1949.

\bibitem{rubin1980simulation}
{Rubin, B.} and {Vinsome, P. K. W.}
\newblock The simulation of the in-situ combustion process in one dimension
  using a highly implicit finite-difference scheme.
\newblock {\em Journal of Canadian Petroleum Technology}, 19(04), 1980.

\bibitem{rubin1985general}
{Rubin, B.} and {Vinsome, P. K. W.}
\newblock A general purpose thermal model.
\newblock {\em Society of Petroleum Engineers Journal}, 25(02):202--214, 1985.

\bibitem{smith1971simulation}
{Smith, J. T.} and {Ali, S. M. }.
\newblock Simulation of in-situ combustion in a two-dimensional system.
\newblock In {\em Fall Meeting of the Society of Petroleum Engineers of AIME}.
  Society of Petroleum Engineers, 1971.

\bibitem{soave1972equilibrium}
{Soave, G.}
\newblock Equilibrium constants from a modified redlich-kwong equation of
  state.
\newblock {\em Chemical Engineering Science}, 27(6):1197--1203, 1972.

\bibitem{stone1970probability}
{Stone, H. L.}
\newblock Probability model for estimating three-phase relative permeability.
\newblock {\em Journal of Petroleum Technology}, 22(02):214--218, 1970.

\bibitem{stone1973estimation}
{Stone, H. L.}
\newblock Estimation of three-phase relative permeability and residual oil
  data.
\newblock {\em Journal of Canadian Petroleum Technology}, 12(4), 1973.

\bibitem{thomas1963study}
{Thomas, G. W.}
\newblock A study of forward combustion in a radial system bounded by permeable
  media.
\newblock {\em Journal of Petroleum Technology}, 15(10):1145--1149, 1963.

\bibitem{ursenbach2010air}
{Ursenbach, M. G.}, {Moore, R. G.}, and {Mehta, S. A.}
\newblock Air injection in heavy oil reservoirs-a process whose time has come
  (again).
\newblock {\em Journal of Canadian Petroleum Technology}, 49(01):48--54, 2010.

\bibitem{van1910equation}
{Van der Waals, J. D.}
\newblock The equation of state for gases and liquids.
\newblock {\em Nobel lectures in Physics}, 1:254--265, 1910.

\bibitem{vaughn1986numerical}
{Vaughn, P.} and {Wyoming, L.}
\newblock {\em A Numerical Model for Thermal Recovery Processes in Tar Sand:
  Description and Application}.
\newblock Morgantown Energy Technology Center, US Department of Energy., 1986.

\bibitem{youngren1980development}
{Youngren, G. K.}
\newblock Development and application of an in-situ combustion reservoir
  simulator.
\newblock {\em Society of Petroleum Engineers Journal}, 20(01):39--51, 1980.

\bibitem{zhu2011efficient}
{Zhu, Z.}
\newblock {\em Efficient Simulation of Thermal Enhanced Oil Recovery
  Processes}.
\newblock PhD thesis, Stanford University, 2011.

\bibitem{zhu2011upscaling}
{Zhu, Z.}
\newblock Upscaling for field-scale in-situ combustion simulation.
\newblock In {\em SPE Annual Technical Conference and Exhibition}. Society of
  Petroleum Engineers, 2011.

\end{thebibliography}
\end{document}